\documentclass[aps,prd,twocolumn,superscriptaddress,nofootinbib,eqsecnumm,showpacs]{revtex4-1}
\usepackage[pass,letterpaper]{geometry}
\pdfoutput=1  
\usepackage[]{graphicx}
\usepackage{color}
\usepackage{latexsym,amsmath,amsfonts,amssymb}
\usepackage[pdfencoding=auto]{hyperref}
\usepackage{hhline}
\usepackage{cancel}
\usepackage{booktabs}
\usepackage{verbatim}
\definecolor{MyBlue}{rgb}{0.15,0.15,0.70}
\hypersetup{
colorlinks=true,
citecolor=MyBlue,
linkcolor=MyBlue,
urlcolor=
}

\newcommand{\be}{\begin{equation}}
\newcommand{\ee}{\end{equation}}
\newcommand{\beq}{\begin{equation}}
\newcommand{\eeq}{\end{equation}}
\newcommand{\bea}{\begin{eqnarray}}
\newcommand{\eea}{\end{eqnarray}}

\newcommand{\dd}{\text{d}}

\newcommand\ees{\end{eqnarray}}
\newcommand\bees{\begin{eqnarray}}

\begin{document}

\title{Fifth force induced by a chameleon field on nested cylinders}

\author{Martin Pernot-Borr\`as}
\email{martin.pernot\_borras@onera.fr}
\affiliation{DPHY, ONERA, Universit\'e Paris Saclay, F-92322 Ch\^atillon, France}
\affiliation{Institut d'Astrophysique de Paris, CNRS UMR 7095,
Universit\'e Pierre \& Marie Curie - Paris VI, 98 bis Bd Arago, 75014 Paris, France}

\author{Joel Berg\'e}
\affiliation{DPHY, ONERA, Universit\'e Paris Saclay, F-92322 Ch\^atillon, France}

\author{Philippe Brax}
\affiliation{Institut de Physique Th\'eorique, Universit\'e Paris-Saclay, CEA, CNRS, F-91191 Gif-sur-Yvette Cedex, France}

\author{Jean-Philippe Uzan}
\affiliation{Institut d'Astrophysique de Paris, CNRS UMR 7095,
Universit\'e Pierre \& Marie Curie - Paris VI, 98 bis Bd Arago, 75014 Paris, France}
\affiliation{Sorbonne Universit\'es, Institut Lagrange de Paris, 98 bis, Bd Arago, 75014 Paris, France}

\date{today}
\begin{abstract}
This article investigates the properties of a scalar fifth force that arises in a scalar tensor-theory with a chameleon screening mechanism in the context of gravity space missions like the MICROSCOPE experiment. In such an experiment, the propagation of the chameleon field inside the nested cylinders of the experiment causes a fifth force when the cylinders are not perfectly co-axial. We propose a semi-analytic method to compute the field distribution and the induced fifth force and compare it to a full numerical simulation, in settings where the cylindrical symmetry is broken. The scaling of the fifth force with both the parameters of the model and the geometry of the experiment is discussed. We show that the fifth force is repulsive, hence adds a destabilizing stiffness that  should  be included in the force budget acting on the detector. This opens the way to a new method to constrain a scalar fifth force in screened models.
\end{abstract}

\maketitle
\tableofcontents

\section{Introduction}

Scalar-tensor theories represent a large class of extensions of General Relativity (GR) that are widely studied~\cite{Will:2018bme,Will:2014kxa,Damour:1993hw} to constrain deviations from GR and to investigate the physical effects of a potential scalar partner to the graviton, which may arise from high-energy theories, e.g. string theory. On large scales, coupled scalars modify the evolution of the universe and its structure. They have attracted a lot of attention in connection with the modeling of the late acceleration of the Universe, i.e. as possible dark energy candidates~\cite{2013arXiv1309.5389J}. On smaller scales, the extra-degree of freedom is responsible for a fifth force. The properties of this fifth force depends on the nature of the couplings of the scalar field to standard matter, universal or not, on the mass of the scalar field, and more generally on its potential. While light field models can be  attracted toward General Relativity~\cite{Damour:1992kf,DAMOUR1994532} and are constrained in  laboratory and space gravity experiments, local tests are more difficult for models exhibiting screening as they require to take into account the effects of the environment~\cite{Berge:2018htm}. Amongst such models, let us cite the symmetron~\cite{PhysRevLett.104.231301} and the chameleon~\cite{khoury_chameleon_2004a,khoury_chameleon_2004} mechanisms. In both cases, the profile of the scalar field and thus the associated fifth force depends on the local mass density: the field acquires a large mass in high density environments responsible for the suppression of the fifth force, whereas in low density environments the force can be long-ranged.

The main goal of this article is to continue our investigation on the possibility to test such scalar-tensor theories with a screening mechanism in gravity space experiments. Even if the coupling of the field is universal, it can generate composition dependent fifth force between macroscopic objects since the profile of the scalar field, and thus the fifth force that derives from it, inside the object depends on its density, and thus on this composition. So far, many experiments~\cite{burrage_tests_2018,BraxReview,Berge:2017ovy}  have set constraints on the existence of a chameleon field among which atom interferometry~\cite{burrageAtom, sabulsky_experiment_2018}, Casimir effect measurements~\cite{brax_detecting_2007} or torsion balance experiments~\cite{upadhye_dark_2012}. Space-borne experiments -- as the MICROSCOPE mission~\cite{touboul_microscope_2017} testing the weak equivalence principle in orbit -- were originally argued to be a possible smoking gun for the chameleon mechanism~\cite{khoury_chameleon_2004a,khoury_chameleon_2004} as the local density in space is much smaller than at the surface of the Earth, hence leading to a lighter field and a stronger fifth force. However, this intuitive argument requires to be analyzed in depth, in particular to take into account the fact that the experimental set-up can itself screens the chameleon. Understanding this screening and the propagation of the scalar field inside the measurement device is a key issue to detect or constrain such  a mechanism. It requires to determine the field  profile for non trivial matter distributions as the theory is highly non linear and a special attention to the boundary conditions must be paid. Multiple approaches have been used  involving both analytic ~\cite{khoury_chameleon_2004, brax_detecting_2007, Brax:2013cfa, PhysRevD.87.105013,  burrage_probing_2015, burrage_proposed_2016, ivanov_exact_2016, nakamura_chameleon_2018, kraiselburd2018, kraiselburd2019} and numerical methodes~\cite{upadhye_dark_2012, hamilton_atom-interferometry_2015, elder_chameleon_2016,schlogel_probing_2016, burrage_shape_2018}.

In a previous paper~\cite{PhysRevD.100.084006}, we considered an idealized experimental setting modeled by  cylindrically or spherically nested geometries, and studied the propagation of a chameleon field inside such a setting. This clarified the occurrence of the screening mechanism and led us to conclude that for experiments similar to the MICROSCOPE mission, the screening induced by the experiment's cavity steps in for most of the parameter space of the chameleon model, hence reducing the hope of constraining  chameleons with this space experiment. Nevertheless, the different parts of the detector are subject to a series of non-gravitational forces that need to be compensated. It follows that the inner cylinders of the device can move and thus depart from the cylindrical symmetry. This can induced an internal source for the fifth force that needs to be modeled and  constrained in  the force budget of the experiment.

To that purpose, we consider a model configuration similar to the MICROSCOPE geometry involving an accelerometer composed of nested test mass cylinders and electrode cylinders. A force on a test mass appears when the cylindrical symmetry is broken by shifting the cylinder from its axis. The goal of this article is to quantify the fifth force induced by this non-coaxiality. Thus, we consider a static configuration of two infinite nested cylinders. After summarizing briefly the theoretical context in Section \ref{sec:SecI}, we start by a simplified exercise in Section~\ref{sec:SecI} in which we restrict to 1-dimensional configurations. Then we tackle the case of nested cylinders by first developing a semi-analytical multipolar expansion in Section \ref{sec:multipolar} and the full numerical integration in Section \ref{sec:fullnum}. Both methods have their own domain of validity and are compared when they both apply. Once the profile are determined, we compute in Section~\ref{sec:force} the resulting force on the inner cylinder and then discuss its scaling with the geometry of the model and the parameters of the theory.

This provides the first analysis of the fifth force stiffness induced by a chameleon field on an idealized gravity experiment with a design similar to the MICROSCOPE mission. It shows that the fifth force being repulsive, it adds a destabilizing stiffness that would require to be compared to the other forces acting on the detector, from electrostatic and Newtonian origin (since the Newtonian force vanishes only for infinite cylinders). Hence, this work paves the way to the analysis of the MICROSCOPE experiment that shall be presented in a companion article~\cite{artsuivant}.

\section{General equations}\label{sec:SecI}

\subsection{Theory}		

Let us consider the theory defined by the general scalar-tensor action in the Einstein frame,
\begin{equation}
\begin{split}
S = \int {\rm d}x^4 ~ &\sqrt{-g} \left[\frac{{\rm M}_{ \rm Pl}^2}{2}{\displaystyle R} -  \frac{1}{2}\partial^\mu\phi\partial_\mu\phi - V(\phi)\right]\\
&- \int {\rm d^4 }x \mathcal{L}_{\rm m}(\widetilde{g}_{\mu\nu}, \phi, ...),
\end{split}
\end{equation}
where $\phi$ is a scalar field, $V$ its potential, ${\rm M}_{ \rm Pl}$ the reduced Planck mass, ${\displaystyle R}$ the Ricci scalar, $g_{\mu\nu}$ the Einstein frame metric, $g$ its determinant and $\mathcal{L}_{\rm m}$ the matter Lagrangian. The field couples non-minimally to matter through the Jordan frame metric
\begin{equation}
\widetilde{g}_{\mu\nu} = A^2(\phi)g_{\mu\nu},
\end{equation}
where $A(\phi)$ is a universal coupling function, from which the dimensionless coupling constant
\begin{equation}
\beta(\phi) = {\rm M}_{ \rm Pl} \frac{{\rm dln}A}{{\rm d}\phi}
\end{equation}
can be defined. It characterizes the magnitude of the coupling to the scalar field to standard matter, and hence the magnitude of the fifth force. Note that the coupling may not be universal, so that the field could have different couplings, $A_i(\phi)$ for the different components of matter. Such models involve  spacetime variations of fundamental constants that have been well-contrained~\cite{Uzan:2002vq,Uzan:2010pm,Uzan:2004qr} so that we restrict our analysis to a universal coupling. The method proposed here generalizes itself easily to non universal couplings.

In the Einstein frame, the scalar field dynamics follows from the Klein-Gordon equation,
\begin{equation}\label{e.KG}
\Box\phi = \frac{{\rm d}V}{{\rm d}\phi} - \frac{\beta(\phi)}{{\rm M}_{\rm Pl}} T_{\mu\nu}g^{\mu\nu},
\end{equation}
so its source term depends both on  the potential and the local value of the trace of the matter stress-energy tensor, which reduces to the local energy density for a non-relativistic matter.

\subsection{Chameleon models}

Chameleon models posits that the potential $V$ and coupling function $A$ do not have the same convexity so that the minimum of the effective potential depends on the local matter density. We shall assume that the coupling function is of the form
\be\label{e.defA}
  A = \hbox{e}^{\beta \phi/{\rm M}_{\rm Pl}}
\ee
and the potential is of the form
\begin{equation}\label{e.defV}
V= \Lambda^4\left(1+\frac{\Lambda^n}{\phi^n}\right)
\end{equation}
where $\Lambda$ is a mass scale, $n$ a natural number and $\beta$ a positive constant. It follows that the Klein-Gordon equation~(\ref{e.KG}) reduces to
\begin{equation}\label{e.kg7}
\Box\phi = \frac{{\rm d}V_{\rm eff}}{{\rm d}\phi}
\end{equation}
with the effective potential
\begin{equation}
 V_{\rm eff}= V(\phi) + \frac{\beta}{{\rm M}_{\rm Pl}} \rho \, \phi,
\end{equation}
$\rho$ being the mass density configuration. This equation enjoys a density-dependent  minimum
\begin{equation}
 \phi_*(\rho)=\left(\frac{{\rm M}_{\rm Pl} n \Lambda^{n+4}}{\beta \rho} \right)^{\frac{1}{n+1}}.
 \label{phi_min}
\end{equation}
In media of constant density $\rho$ the field would tend to reach this minimal value. This would occur on scales given by the density dependent Compton wavelength,
\begin{equation}
\lambda_{\rm c}(\rho_{\rm}) \equiv \sqrt{\frac{1}{n (n + 1) \, \Lambda^{n+4}} \left(\frac{n \,{\rm M}_{\rm Pl} \, \Lambda^{n+4}}{\beta \, \rho_{\rm}}\right)^\frac{n+2}{n+1}},
\label{Comp_wl}
\end{equation}
which becomes shorter as $\rho$  is  larger.

Finally, if we assume static configurations, the field is governed by the Laplace equation
\begin{equation}\label{e.KG3}
 \Delta\phi = n\Lambda^{n+4}\left[\phi_*^{-(n+1)} - \phi^{-(n+1)}\right)].
\end{equation}
With the rescalings $\phi/\Lambda\rightarrow\phi$, $\beta/\Lambda^3 \rightarrow \beta$ and $\Lambda {\bf r} \rightarrow {\bf r}$, it reduces to
\begin{equation}\label{e.KG-gen}
 \Delta\phi = \frac{\beta}{{\rm M}_{\rm Pl}}\rho({x})- n \phi^{-(n+1)}.
\end{equation}
From such a rescaling, a profile computed for specific $\Lambda$, $\beta$ and matter configurations specified by  $\rho$, could directly lead to profiles for different $\Lambda$ and $\beta$ and a rescaled geometry.

\section{One-dimensional asymmetric configurations}\label{sec:1D}

Let us first start by considering a non-symmetrical one-dimensional model. It consists of 3 infinite parallel walls of the same thickness. The central wall can move and thus is not necessarily at the same distance from the other two external fixed walls.

The configuration is characterized by the thickness $e$ of the walls, the distance $2g+e$ between the two external walls and the displacement $\delta$ of the central wall with respect to the middle position. The density of the walls and of the inter-wall regions are respectively denoted by  $\rho_{\rm in}$ and $\rho_{\rm vac}$.  Throughout this work, if not stated otherwise, we shall assume
$$
\rho_{\rm in} = 8.125 \,{\rm g.cm}^{-3},
$$
 and
$$
\rho_{\rm vac} = 10^{-3} \rho_{\rm in}
$$
for which the corresponding Compton wavelengths are $\lambda_{\rm c, in} \simeq 2 \,{\rm cm}$ and $\lambda_{\rm c, vac} \simeq 2 \,{\rm m}$.

\subsection{Resolution method}

The profile of the field in the symmetrical case ($\delta = 0$) has already been described in our former work~\cite{PhysRevD.100.084006}. We can adapt the method to deal with the non-symmetrical case and solve Eq.~\eqref{e.KG-gen} for a non-symmetrical configuration.

The main problem is to determine the boundary conditions for the numerical integration. When $\delta = 0$, it is obvious, by symmetry, that the field's derivative cancels at the center of the central wall. Under that condition, one can proceed by  dichotomy on the value of the field at the center to determine the value that is compatible with  the boundary conditions at  large distance.

When $\delta \neq 0$, the derivative of the field does not vanish at the center but at  a slightly shifted location that depends on $\delta$. Again, it can be determined by dichotomy. Since the central wall is separated from  the two external walls by distances of  respectively $g-\delta$ and $g+\delta$, we start at some initial position $x_0$ in the central wall. We then determine the profile that corresponds to the condition $\phi'(x_0) = 0$ with the same procedure as for the symmetrical case. The different configurations encountered in each direction -- i.e. a gap of respective width $g-\delta$ and $g+\delta$ -- and the boundary conditions, give two different values of $\phi(x_0)$. Depending on the sign of the difference of these values we adjust $x_0$, and repeat the procedure until convergence when this difference gets negligible. This way we obtain the correct position $x_0$ and initial value $\phi(x_0)$ corresponding to the profile satisfying the correct boundary conditions at large distance.

\subsection{Profile of the field and resulting force on the central wall}

\begin{figure}
\includegraphics[width = 1\columnwidth]{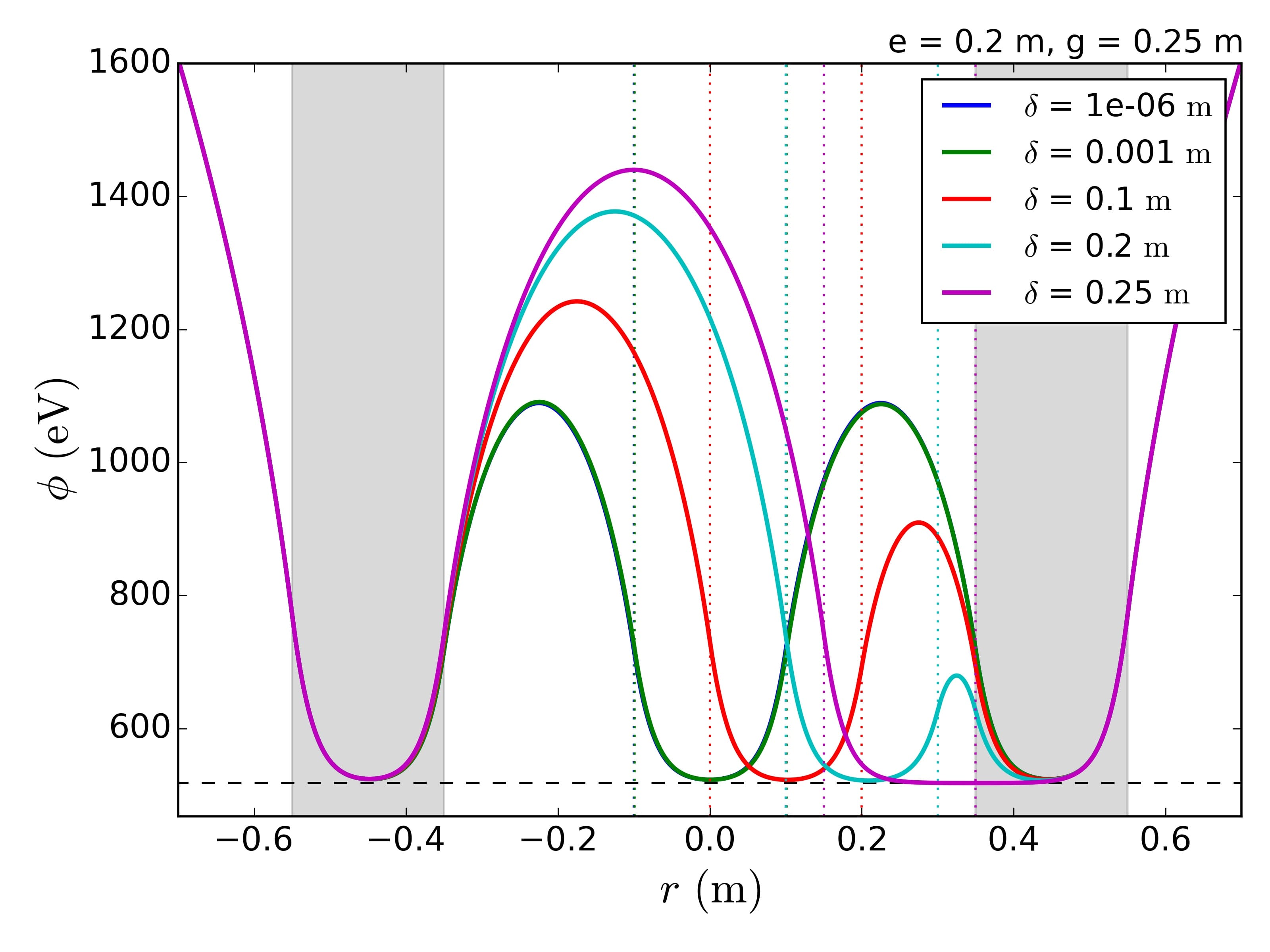}
\caption{Profiles of $\phi$ for a three-wall asymmetric configuration for walls of thickness $e=0.2 \,{\rm m}$ and for different displacements $\delta$ of the central wall. The doted lines delimit the borders of the central walls. The shaded zones correspond to the two fixed external walls. The model parameters have been chosen to $n=2$, $\beta = 1$, $\Lambda = 1 \, {\rm eV}$.}
\label{figphi3walls}
\end{figure}

\begin{figure}
\includegraphics[width = 1\columnwidth]{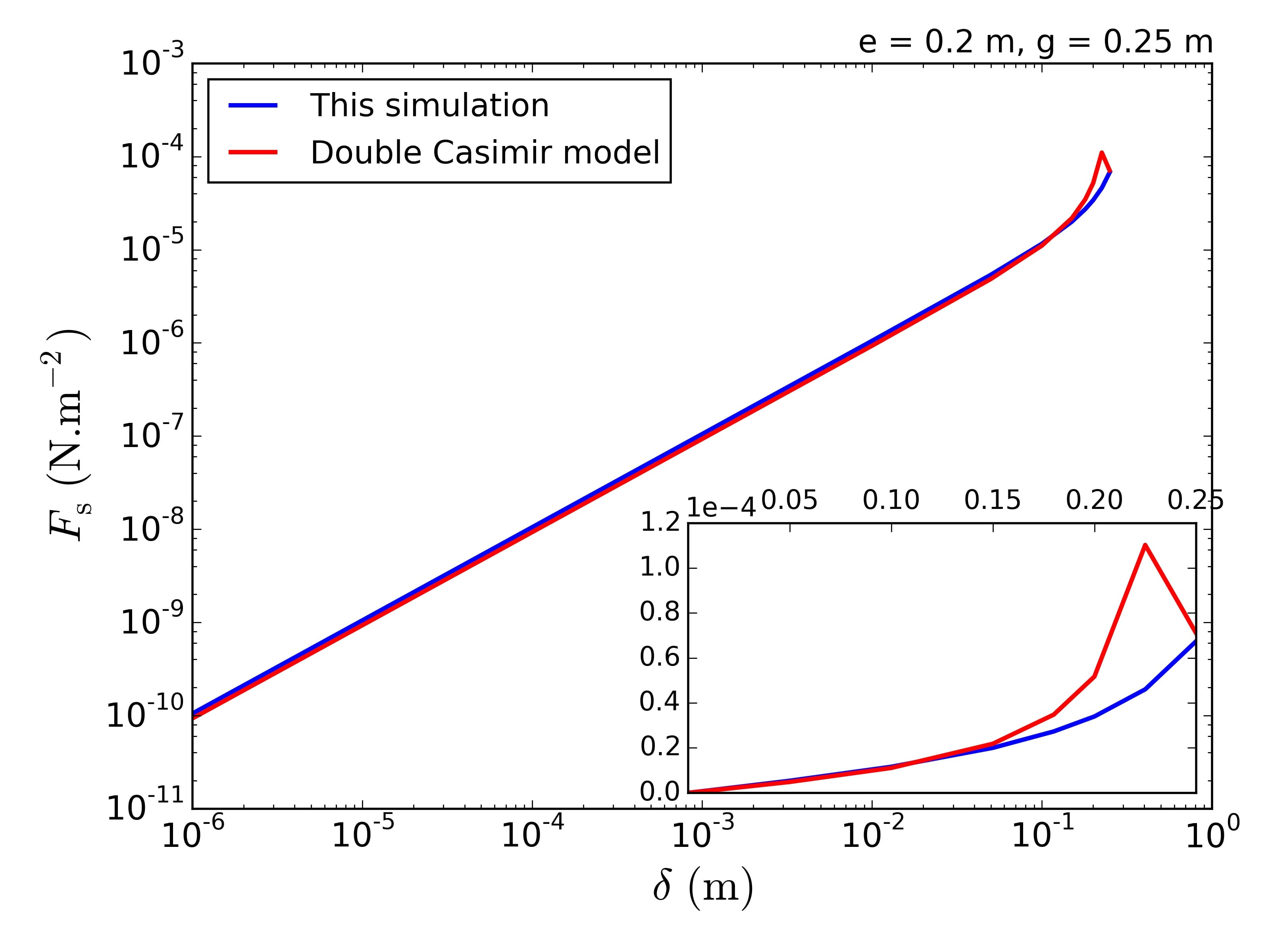}
\caption{The pressure on the central wall for the configuration described in Fig.~\ref{figphi3walls} as a function of the displacement $\delta$.  The red line corresponds to the two-Casimir-like configuration as computed in Ref.~\citep{brax_detecting_2007}.}
\label{figforce3walls}
\end{figure}

The profile of the field for a configuration in which  $e = 0.2 \,{\rm m}$ and $g = 0.25 \,{\rm m}$ is depicted in Fig.~\ref{figphi3walls} for different displacement $\delta$. Since the profile is no more symmetrical inside the central wall, it implies that the integration of the fifth force $-\beta\nabla\phi/M_{\rm Pl}$ does not vanish. Figure \ref{figforce3walls} depicts the evolution of the fifth force with $\delta$. For small displacements it is linear with a positive sign, i.e. a repulsive force that tends to destabilize the configuration. It develops a non-linear scaling for large $\delta$. Note, for comparison, that the Newtonian force on the central wall remains zero whatever $\delta$.

This result can be compared to the one obtained by considering this problem as two joined Casimir-like configurations -- two sets of parallel plates whose chameleonic force has been analytically computed in Ref.~\cite{brax_detecting_2007} -- where the central wall is pulled by each external walls resulting in a total destabilizing force. The agreement between our numerical computation and this analytical form is excellent for almost all the range of displacements. The Casimir-like force scales for one pair of plates as $d^{-1}$ for $n=2$. Then applied to our cases here it scales as $\frac{2 \delta}{(g+\delta)(g-\delta)}$. So it is linear as long $\delta\ll g$. For larger displacement the agreement is not as good since our result departs from linearity for larger $\delta$. This is indeed not surprising as for small displacements this is the  regime where $\lambda_{c, in}\ll d \ll\lambda_{c, vac}$ for both sets of plates, for which  a good agreement already exists \cite{brax_detecting_2007}. For large displacements this is no longer the case, explaining the discrepancy.

\section{Two-dimensional cylindrical asymmetric configuration: semi-analytic multipolar approximation}\label{sec:multipolar}

\begin{figure}[htb]
\centering
 \includegraphics[width = 1\columnwidth]{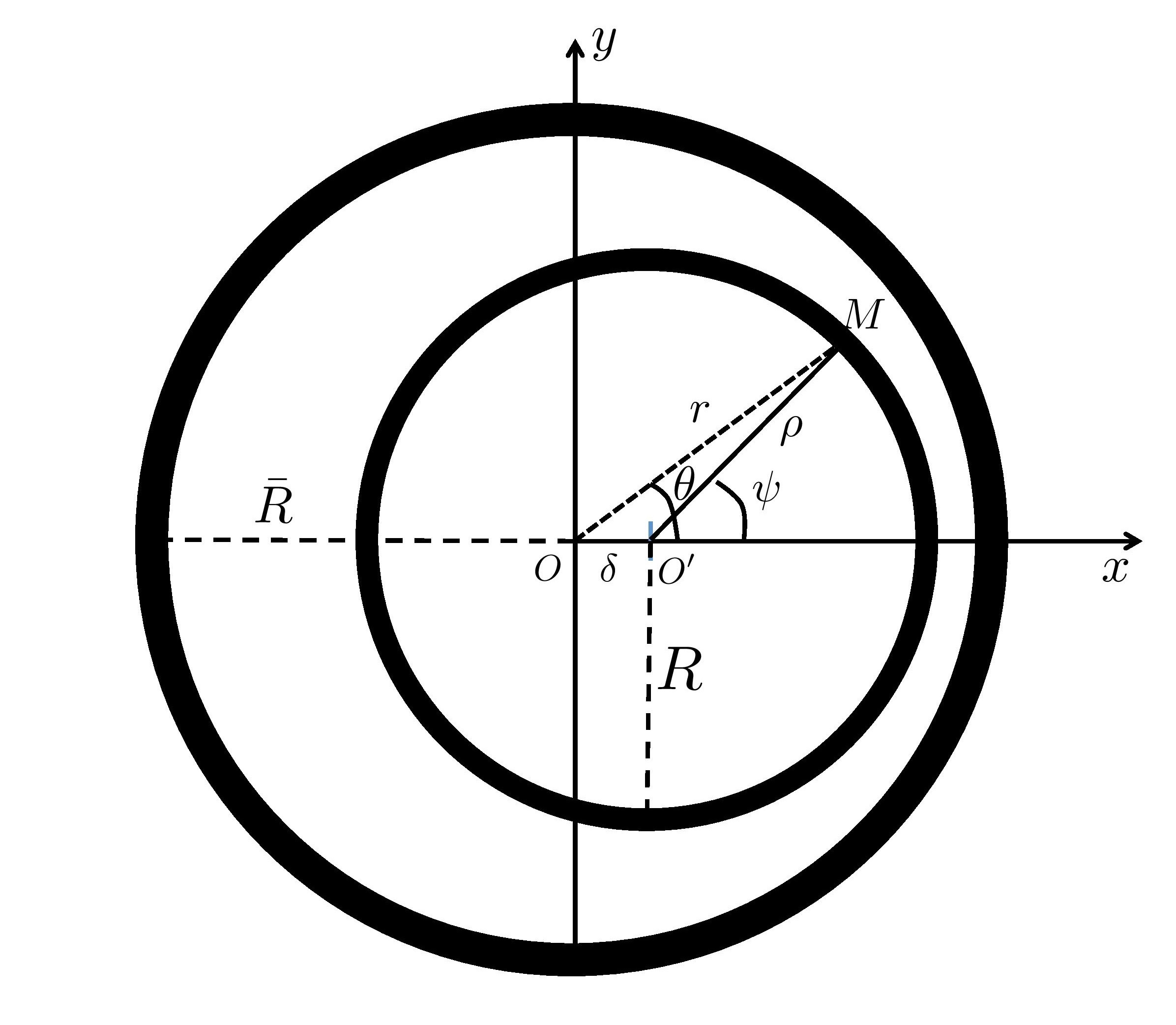}
   \caption{Geometry of the 2-dimensional configuration of the two nested cylinders and definition of the notations of the problem.}
   \label{Fig-Cyl1}
\end{figure}

Let us now turn to the less academic case of two infinite nested cylinders. This geometry is close to the one of MICROSCOPE's accelerometers even though we still assume that the cylinders are infinite to simplify the analysis.  The transverse geometry is detailed in Fig.~\ref{Fig-Cyl1}  and the goal is to compute the force on the inner cylinder once shifted from the center. This is indeed a more complex problem than previously as it requires to treat the full 2 dimensions in Eq.~\eqref{e.KG-gen} and cannot be reduced to 1-dimensional problem as for configurations with cylindrical symmetry. Nevertheless, as we shall now see, for small displacements the problem can be simplified using a multipolar expansion of the field configuration.

The geometry we consider is described on Fig.~\ref{Fig-Cyl1} and consists of two cylinders:
\begin{itemize}
\item an outer cylinder of radius $\bar R$ and width $\bar e$ centered on $O$ and with density $\rho_{\rm out}$;
\item an inner cylinder of radius $R$ and width $e$ centered on $O'$ and with density $\rho_{\rm in}$. We assume that
\begin{equation}
 \bf{OO'} = \delta \, {\bf e}_x,
\end{equation}
where $\delta$ is the displacement of the inner cylinder with respect to the axis of symmetry and $ {\bf e}_x$ the unit vector in this direction, arbitrarily chosen to be the $x$-axis.
\end{itemize}
We define the basis of the Cartesian coordinates as $({\bf e}_x,{\bf e}_y)$ and the polar coordinates system $({\bf e}_r,{\bf e}_\theta)$ with
 \begin{equation}
   {\bf e}_x.{\bf e}_r=\cos\theta, \quad
    {\bf e}_y.{\bf e}_r=\sin\theta,
 \end{equation}
and
 \begin{equation}
   {\bf e}_x.{\bf e}_\theta = -\sin\theta,\quad
   {\bf e}_y.{\bf e}_\theta =  \cos\theta.
 \end{equation}

In complex notations, it is clear that the equation of the inner cylinder is $r(\theta) \hbox{e}^{i\theta}=\delta + R \hbox{e}^{i\psi}$ so that $R^2=r^2+\delta^2-2\delta r \cos\theta$, from which we determine the equation of the inner disk in polar coordinates
\begin{equation}\label{e.incoord}
 r(\theta) = R\left(\frac{\delta}{R}\cos\theta +\sqrt{1-\frac{\delta^2}{R^2}\sin^2\theta}  \right)
\end{equation}
from which we can define the inner and outer borders of the inner cylinder as $r_-(\theta)$ and $r_+(\theta)$, respectively with $R$ and $R+e$ in Eq.~(\ref{e.incoord}). It follows that the matter density is distributed as
\begin{equation}
 \rho(r,\theta) = \rho_{\rm vac} +
  \left\lbrace
  \begin{array}{lcl}
   \rho_{\rm in}-\rho_{\rm vac} & \hbox{if} & r \in [r_-(\theta),r_+(\theta)] \\
   \rho_{\rm out}-\rho_{\rm vac} & \hbox{if}  & r \in [\bar R, \bar R + \bar e]  \\
   0 && \hbox{otherwise}\\
 \end{array}
 \right. .
\end{equation}
It will turn  convenient to define the function $\Xi$ such that $\Xi(x;a,b)=1$ is $x\in[a,b]$ and 0 otherwise, {\em i.e.} it is defined in terms of the Heaviside distribution $H$ as
\begin{equation}\label{e.dfXi}
\Xi(x;a,b) = H(x-a)- H(x-b).
\end{equation}
It follows that
\begin{eqnarray}\label{e.rho18}
\rho(r,\theta) &=& \rho_{\rm vac} +( \rho_{\rm out}-\rho_{\rm vac} )\Xi(r;\bar R, \bar R+\bar e) \nonumber\\
 && \qquad + (\rho_{\rm in}-\rho_{\rm vac})\Xi[r;r_-(\theta),r_+(\theta)].
\end{eqnarray}

\subsection{Mode decomposition}

In cylindrical coordinates, forgetting about the $z$-dimension since by symmetry $\phi(r,\theta)$, the gradient is given by $\nabla=(\partial_r,\partial_\theta/r)$ and the Laplacian by
\begin{equation}
 \Delta f= \partial_r^2 f + \frac{1}{r}\partial_r f+ \frac{1}{r^2}\partial_\theta^2 f.
\end{equation}
for any function $f(r,\theta)$. One can always decompose $f$ in modes as
\begin{equation}\label{e.md}
 f(r,\theta) = \sum_{\ell\in Z} \frac{u_\ell(r)}{\sqrt{r}}\hbox{e}^{i\ell \theta}
\end{equation}
with
\begin{equation}
  \frac{u_\ell(r)}{\sqrt{r}}=\int\frac{{\rm d}\theta}{2\pi} f(r,\theta)\hbox{e}^{-i\ell \theta}.
\end{equation}
It follows that
\begin{equation}
 \Delta f = \frac{1}{\sqrt{r}} \sum_{\ell\in Z}  \left[ u_\ell'' +\frac{\left(\frac{1}{4}-\ell^2\right)}{r^2} u_\ell \right]\hbox{e}^{i\ell \theta}.
\end{equation}

Let us now turn to integration. We will have to integrate functions $f(r,\theta)$, such as the components of the force, on the inner cylinder as
$$
\int f(M) {\rm d}m =
\rho_{\rm in} h \int_{\hbox{inner cyl.}} f(r,\theta) r{\rm d}\theta {\rm d}r,
$$
$h$ being the length of the cylinder. For each $\theta$, $r$ varies between $r_-$ and $r_+$ so that
\begin{equation}
 \int f(M) {\rm d}m =  \rho_{\rm in} h \int_0^{2\pi} {\rm d}\theta \int_{r_-(\theta)}^{r_+(\theta)}   f(r,\theta) r {\rm d}r.
\end{equation}
It is then ``easily'' checked that for $f=1$ we get the mass of the cylinder $\rho_{\rm in}\pi h e(2R+e)$. Indeed this is a tricky integral which turns out to be trivial in terms of the angle $\psi$ defined in Fig.~\ref{Fig-Cyl1}.

\subsection{One cylinder}

We start by considering only the outer cylinder. The density profile has been fully described in our former work~\cite{PhysRevD.100.084006} and is denoted by $\bar \phi(r)$. It is solution of
\begin{equation}\label{e.2cyl-bar}
 \bar\phi''+\frac{1}{r}\bar\phi' = \frac{\beta}{{{\rm M}_{\rm Pl}}}\left[\rho_{\rm vac} +( \rho_{\rm out}-\rho_{\rm vac} )\Xi(r;\bar R, \bar R+\bar e) \right]- \frac{n}{\bar\phi^{n+1}}
\end{equation}
with the boundary conditions
\begin{equation}
 \bar\phi(\infty)=\phi_*(\rho_{\rm vac}), \qquad
 \bar\phi'(0)=0.
 \label{phibarcond}
\end{equation}

\subsection{Two cylinder configuration}

Starting from the previous profile $\bar \phi(r)$, we consider the effect of the second cylinder and decompose $\phi$ as
\begin{equation}
 \phi(r,\theta) = \bar\phi(r) + \psi(r,\theta).
\end{equation}
Indeed, if the inner cylinder is centered in $O$ then $\psi$ in only a function of $r$. Such configurations were also studied in our previous work~\cite{PhysRevD.100.084006}. Now by subtracting Eq.~(\ref{e.2cyl-bar}) to the Klein-Gordon equation~(\ref{e.KG-gen}) we get
\begin{eqnarray}\label{e.2cyl-psi}
  \psi''+\frac{1}{r}\psi' + \frac{1}{r^2}\partial^2_\theta \psi  &=& \frac{\beta}{{{\rm M}_{\rm Pl}}}( \rho_{\rm in}-\rho_{\rm vac} )\Xi[r;r_-(\theta), r_+(\theta)] \nonumber \\
    &+&\frac{n}{\bar\phi^{n+1}(r)} - \frac{n}{[\bar\phi(r)+\psi(r,\theta)]^{n+1}}.
\end{eqnarray}
This equation is fully general and no approximation has been made so far so far. It is a 2-dimensional  non-linear partial differential equation. There is no way it can be analytically solved in  full generality.

\subsection{Multipolar hierarchy}

To go further, we decompose $\psi$ in multipoles as in Eq.~(\ref{e.md}) and we single out the monopole $\ell=0$,
\begin{equation}\label{e.decpsi}
 \psi(r,\theta) = \psi_0(r) + \sqrt{\frac{\delta}{r}}\sum_{\ell\not=0} u_\ell(r)\hbox{e}^{i\ell \theta}.
\end{equation}
This decomposition is fully general. Since $\psi$ is a real-valued function, $u_\ell^*=u_{-\ell}$. We introduce the dimensionless factor $\delta/r$ as it is clear that the non-radial terms all vanish when $\delta=0$ and that $\delta/R\sim\delta/(R+e)$ will serve as a small parameter for our expansion. Thus, the generic Klein-Gordon equation takes the form
\begin{eqnarray}\label{e.2cyl-ul}
 && \psi_0''+\frac{1}{r}\psi_0' + {\sqrt\frac{\delta}{r}} \sum_{\ell\not=0}  \left[ u_\ell'' +\frac{\left(\frac{1}{4}-\ell^2\right)}{r^2}  u_\ell \right]\hbox{e}^{i\ell \theta}  = \nonumber \\
  && \qquad\qquad \frac{\beta}{{{\rm M}_{\rm Pl}}}( \rho_{\rm in}-\rho_{\rm vac} )\Xi[r;r_-(\theta), r_+(\theta)] \nonumber \\
    &&\qquad\qquad +\frac{n}{\bar\phi^{n+1}(r)} - \frac{n}{[\bar\phi(r)+\psi(r,\theta)]^{n+1}}.
\end{eqnarray}
The goal is thus to determine the functions $\psi_0(r)$ and $u_\ell(r)$. It is clearly a difficult task as the last term of the r.h.s. couples to all the modes.

\begin{widetext}
The evolution of each mode can be obtained by integrating Eq.~(\ref{e.2cyl-ul}) times $\hbox{e}^{-i\ell'\theta}{\rm d}\theta/2\pi$ over $\theta$ and singling out the monopole from the $\ell\not=0$ modes so that Eq.~(\ref{e.decpsi}) splits as
\begin{eqnarray}
 \psi_0''+\frac{1}{r}\psi_0' &=& \frac{n}{\bar\phi^{n+1}(r)} -\int
  \frac{n}{\left[\bar\phi(r)+\psi_0(r) + \sqrt{\frac{\delta}{r}}\sum_{\ell'\not=0} u_{\ell'}(r)\hbox{e}^{i\ell' \theta}\right]^{n+1}}
  \frac{{\rm d}\theta}{2\pi}\nonumber\\
&&
 \qquad\qquad\qquad\qquad+\frac{\beta}{{{\rm M}_{\rm Pl}}}( \rho_{\rm in}-\rho_{\rm vac} )  \left\lbrace
 \begin{array}{l r l}
 \Xi[r;R, R+e]   & \hbox{if} & \delta = 0\\
 &&\\
  \int \Xi[r;r_-(\theta), r_+(\theta)] \frac{{\rm d}\theta}{2\pi} & \hbox{if} & \delta \not= 0\\
 \end{array}
 \right. \label{e.2cyl-hierarchy0} \\
 {\sqrt\frac{\delta}{r}}\left[ u_\ell'' +\frac{\left(\frac{1}{4}-\ell^2\right)}{r^2}  u_\ell \right] &=&
  \frac{\beta}{{{\rm M}_{\rm Pl}}p}( \rho_{\rm in}-\rho_{\rm vac} )\int \Xi[r;r_-(\theta), r_+(\theta)] \hbox{e}^{-i\ell\theta}\frac{{\rm d}\theta}{2\pi}\nonumber\\
  && \qquad - n\int
  \frac{\hbox{e}^{-i\ell\theta}}{\left[\bar\phi(r)+\psi_0(r) + \sqrt{\frac{\delta}{r}}\sum_{\ell'\not=0} u_{\ell'}(r)\hbox{e}^{i\ell' \theta}\right]^{n+1}}
  \frac{{\rm d}\theta}{2\pi} \label{e.2cyl-hierarchyell}
\end{eqnarray}
Let us note that (1) this hierarchy is highly non-linear and that (2) the complex integrals on the r.h.s. of Eqs.~(\ref{e.2cyl-hierarchy0}-\ref{e.2cyl-hierarchyell}) cannot be performed as one would need to know the poles of its integrand, which depend on the whole solution and because, due to the displacement, the radial width of the inner cylinder depends on $\theta$. Nevertheless as shown in Appendix~A, the integral of $\Xi  \hbox{e}^{-i\ell\theta}$ over $\theta$ can be computed analytically so that the only big issue is the complex integral involving $u_\ell$.
\end{widetext}

\subsection{Small displacement approximation}

So far, the system~(\ref{e.2cyl-hierarchy0}-\ref{e.2cyl-hierarchyell}) is fully general since we made no approximation.  Now, keeping in mind our goal, we want the force on the inner cylinder, so that we are interested on the field configuration on the cylinder, that is close to $r\sim R$. Since we assume $\delta\ll R$, we can expand our solutions in $e/R$.

First, we define $e(\theta)$ as
\begin{equation}
 e(\theta) = r_+(\theta) - r_-(\theta)
\end{equation}
with the definition~(\ref{e.incoord}). At lowest order in $\delta/R$, it reduces to
\begin{equation}
 e(\theta) = e\left(1+\frac{\delta^2}{R(R+e)}\sin^2\theta\right).
\end{equation}
Then, consider Eq.~(\ref{e.2cyl-hierarchy0}).The computation of the integral of $\Xi$ is obtained by taking the limit $\ell\rightarrow0$ in Eq.~(\ref{e.compXi}) as $[\vartheta_+(r) -\vartheta_-(r)]/\pi$ where $\vartheta_+(r)$ and $\vartheta_-(r)$ are two angles in $[0,\pi]$  at which the circle of radius $r$ centered on $O$ intersects the circle centered on $O'$ of radius $R$ and $R+e$ respectively. They are defined only for $r\in[R-\delta,R+\delta]$ and $r\in[R+e-\delta,R+e+\delta] $ respectively so that this term vanishes outside of the support $[R-\delta, R+E+\delta]$. It can be checked that in the limit $\delta\rightarrow0$ it reduces to the function equal to 1 on this support, that is precisely $\Xi[r;R,R+e]$. So, we get
\begin{eqnarray}
 \frac{\beta}{{\rm M}_{\rm Pl}}( \rho_{\rm in}-\rho_{\rm vac} )  \left\lbrace
 \begin{array}{l r l}
 \Xi[r;R, R+e]   & \hbox{if} & \delta = 0\\
 &&\\
  \frac{\vartheta_{+}(r) - \vartheta_{-}(r) }{\pi} & \hbox{if} & \delta \not= 0\\
 \end{array}
 \right.
 \label{e.2cyl-hierarchy0b}
 \end{eqnarray}
for the source term.

Then, consider Eq.~(\ref{e.2cyl-hierarchyell}).  The multipolar components of $\Xi$ are derived in Appendix~A, see Eq.~(\ref{e.compXi}).

Now, we need to treat the non-linear term. To that purpose we consider an expansion in powers of $\delta/r$. The dominant term involves only functions of $r$ so that the integral over $\theta$ vanishes. It follows that
$$\int \hbox{e}^{-i\ell\theta}\left[\bar\phi(r)+\psi_0(r) + \sqrt{\frac{\delta}{r}}\sum_{\ell'\not=0} u_{\ell'}(r)\hbox{e}^{i\ell' \theta}\right]^{-(n+1)} \frac{{\rm d}\theta}{2\pi}
$$
reduces to
$$
 -(n+1)\sqrt{\frac{\delta}{r}}\frac{u_\ell(r)}{\left[\bar\phi(r)+\psi_0(r)\right]^{n+2}}
$$
at lowest order. Then, the first non-linear term is given by
\begin{eqnarray}
 \frac{(n+1)(n+2)}{2}\frac{\delta}{r}\frac{\sum_{L \neq 0} u_L(r)u_{\ell-L}(r)\biggr\rvert_{\ell-L \neq 0}}{\left[\bar\phi(r)+\psi_0(r)\right]^{n+3}}.
\end{eqnarray}
In the equation for $\psi_0$ we have the contribution of the monopole  $-n/[\bar\phi(r)+\psi_0(r)]^{n+1}$ and then the linear term in $u_\ell$ vanishes so that the first correction is the non-linear term involving the sum $\sum_{L \neq 0} u_L(r)u_{-L}(r) = \sum \vert u_\ell(r)\vert^2$.

In conclusion, we get the hierarchy for the modes as a set of 1-dimensional differential equations to which we need to add the equation for $\bar\phi$, so that the full system is described by
\begin{widetext}
\begin{eqnarray}
\bar\phi''+\frac{1}{r}\bar\phi' &=& \frac{\beta}{{\rm M}_{\rm Pl}}\left[\rho_{\rm vac} +( \rho_{\rm out}-\rho_{\rm vac} )\Xi(r;\bar R, \bar R+\bar e) \right]- \frac{n}{\bar\phi^{n+1}}\label{e.2cyl-hierarchyBarFin}\\
 \psi_0''+\frac{1}{r}\psi_0' &=& \frac{n}{\bar\phi^{n+1}(r)}
- \frac{n}{[\bar\phi(r)+\psi_0(r)]^{n+1}}
+\frac{n(n+1)(n+2)}{2}\frac{\delta}{r}\frac{\sum_{L \neq 0} \vert u_L(r)\vert^2}{\left[\bar\phi(r)+\psi_0(r)\right]^{n+3}}\nonumber\\
&&\qquad\qquad+   \frac{\beta}{{\rm M}_{\rm Pl}}( \rho_{\rm in}-\rho_{\rm vac} )  \left\lbrace
 \begin{array}{l r l}
 \Xi[r;R, R+e]   & \hbox{if} & \delta = 0\\
 &&\\
  \frac{\vartheta_{+}(r) - \vartheta_{-}(r) }{\pi} & \hbox{if} & \delta \not= 0\\
 \end{array}.
 \right. \label{e.2cyl-hierarchy0Fin} \\
 \ u_\ell'' +\frac{\left(\frac{1}{4}-\ell^2\right)}{r^2}  u_\ell  &=&
  \frac{\beta}{{\rm M}_{\rm Pl}}( \rho_{\rm in}-\rho_{\rm vac} ) \sqrt{\frac{r}{\delta}}   \frac{\left[\sin\ell\vartheta_{+}(r)- \sin\ell\vartheta_{-}(r)\right]}{\pi\ell}
  \nonumber\\
   && \qquad +n(n+1)\frac{u_\ell(r)}{\left[\bar\phi(r)+\psi_0(r)\right]^{n+2}}
  -\frac{n(n+1)(n+2)}{2}\sqrt{\frac{\delta}{r}}\frac{\sum_{L \neq 0} u_L(r)u_{\ell-L}(r)\biggr\rvert_{\ell-L \neq 0}}{\left[\bar\phi(r)+\psi_0(r)\right]^{n+3}}
  \label{e.2cyl-hierarchyellFin},
\end{eqnarray}
where $\Xi(r;\bar R, \bar R+\bar e)$ is defined in Eq.~(\ref{e.dfXi}), $\vartheta_\pm(r)$ in Eq.~(\ref{e.compVarT}). The equation for $\bar\phi$ is closed and can be solved easily numerically following the same method as in our previous work~\cite{PhysRevD.100.084006}. Then, the equation for $\psi_0$ is coupled to all the modes. But, if we restrict to ${\cal O}(\delta/R)$ it becomes closed. Then, the infinite  set of equations for the $u_\ell$ becomes again linear if we work at order ${\cal O}(\sqrt{\delta/R})$, and we can solve it having previously solved for $\psi_0$. Note that this set of equations is only valid for $\delta<e/2$.
\end{widetext}

\subsection{Numerical scheme}

To completely specified the system, we need to define properly the boundary conditions for $(\psi_0,u_\ell)$.

The total field $\phi$ must verify the same asymptotic boundary condition than $\bar\phi$ : $\phi(\infty)=\phi_*(\rho_{\rm vac})$. Consequently, both the monopole and the multipoles must asymptotically cancel,
\be
\psi_0(\infty) = 0,\qquad
u_{\ell}(\infty) =0.
\ee

We now have all the elements to integrate numerically the set of equations~(\ref{e.2cyl-hierarchyBarFin}-\ref{e.2cyl-hierarchyellFin}). In the following all numerical examples will assume, if not specified otherwise, that the cylinders are of same density $\rho_{\rm in}$ and that the parameters of the geometry are $R = 0.2\,{\rm m}$, $e = 0.05\,{\rm m}$ and $\bar R = 0.6\,{\rm m}$, $\bar e = 0.1\,{\rm m}$.

\subsubsection{Monopole $\psi_0$}

The contribution of the monopole being cylindrically symmetric, its derivative shall cancel at $r=0$ : $\psi_0'(0) = 0$. We can therefore follow the same numerical resolution scheme as we performed for $\bar\phi$ in Ref.~\cite{PhysRevD.100.084006}.

\begin{figure}
\includegraphics[width = 1\columnwidth]{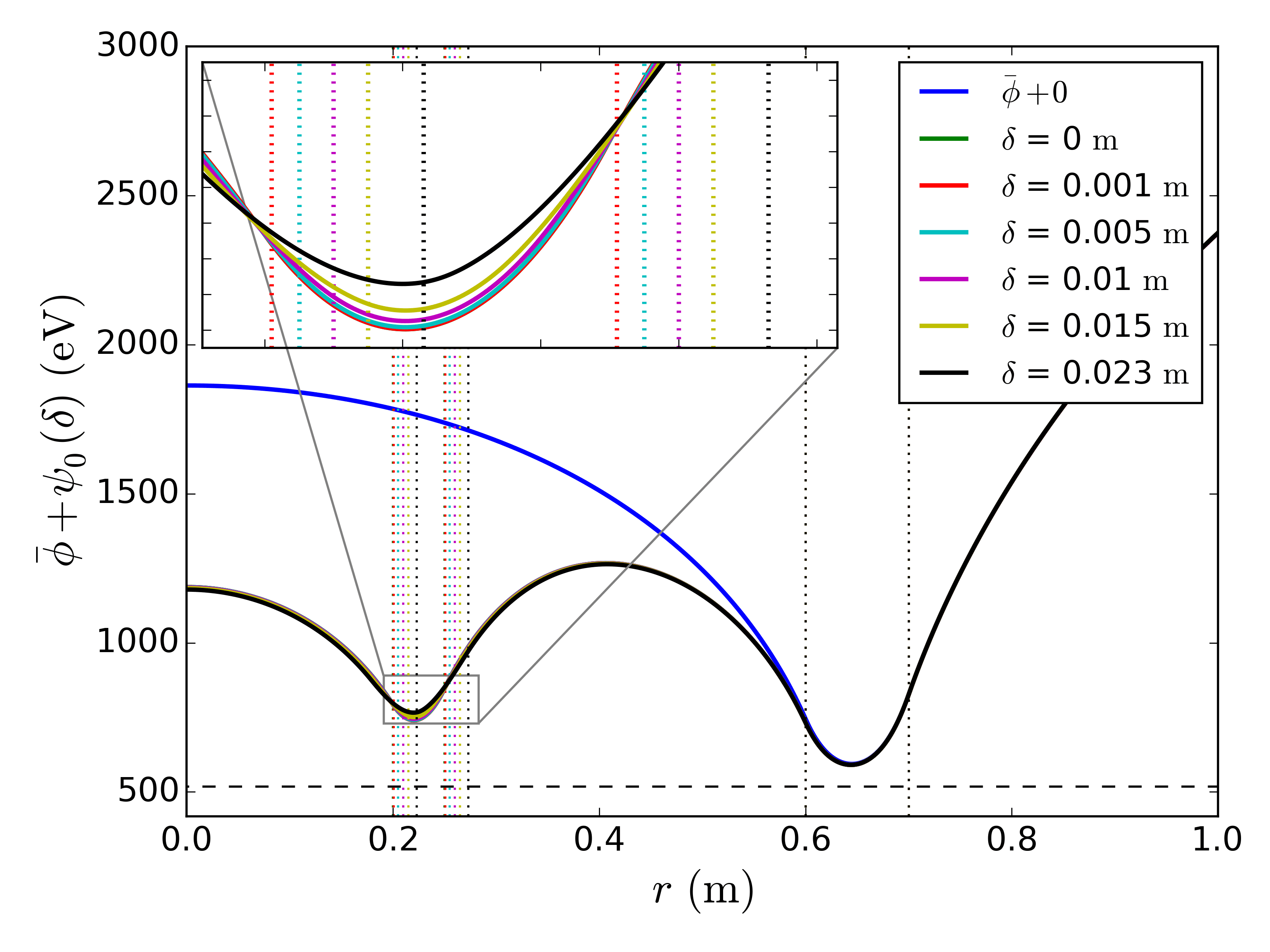}
\caption{Profiles of the field including the monopole correction for an asymmetric system of  two nested cylinders for different displacements $\delta$ of the inner cylinder. The blue line shows the one-cylinder profile $\bar\phi$. The green line is the centered two-nested-cylinder profile. The dotted lines delimit the border of the cylinders.}
\label{fig_psi0}
\end{figure}

Figure \ref{fig_psi0} shows the profile of the monopole for various values of $\delta$. It is compared to the one-cylinder profile $\bar\phi$ and to the symmetrical two-cylinder profile. As expected, it can be checked that the monopole profile tends to the former profile when $\delta$ tends to 0. As $\delta$ gets larger, the minimum value of the field reached in the inner cylinder departs slightly from the corresponding value in the two-centered-cylinders case. The total field might then leak in the multipoles.

\subsubsection{Multipoles $u_\ell$}
\label{sec:sec_multipole}
\begin{figure}
\includegraphics[width = 1\columnwidth]{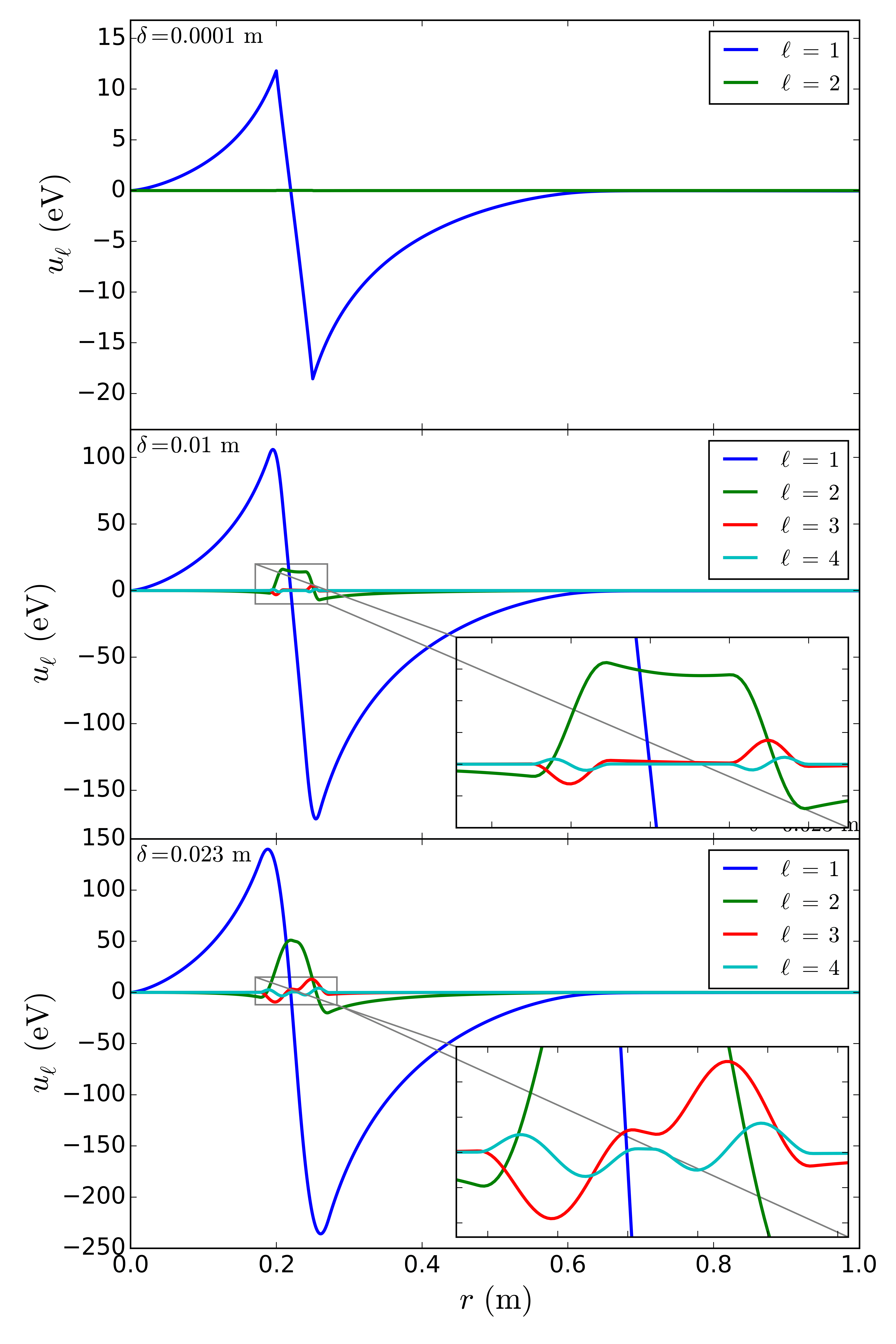}
\caption{Multipoles of order $\ell$ obtained for a set of displacement $\delta = 0.0001, \,0.01, \,0.023 \, {\rm m}$ from top to bottom.}
\label{fig_ul}
\end{figure}

The integration of the multipoles is more complex. Indeed, we do not know the position at which the field's derivative cancels, position used previously as a starting point to integrate $\bar\phi$ and $\psi_0$. Nevertheless Eq.~\eqref{e.decpsi} gives useful information. The factor in front of the multipole sum scales as $1/\sqrt{r}$. For the total field not to diverge at $r=0$, each $u_{\ell}$ must then scale at least as $r^{\frac{1}{2}}$  at $r=0$. We thus deduce that we must have for all $\ell$ : $u_{\ell}(0) = 0$. Similarly to the method used to integrate $\bar\phi$ and $\psi_0$, this leaves us with one parameter $u'_{\ell}(0)$ for the dichotomy which determines the correct initial condition giving the proper profile that verifies $u_{\ell}(\infty) = 0$.

Figure~\ref{fig_ul} depicts the first multipoles for  several displacements of the inner cylinder $\delta$. We observe that, as expected, the contribution of the multipoles is more important for large $\delta$. We also notice that for small $\delta$ the dipole ($\ell = 1$) is the main contribution whereas for larger $\delta$, the $\ell = 4$ term still provides a contribution to the field. We will see in Sec.~\ref{sec:force} that this hierarchy is preserved when computing the force on the inner cylinder, such that the contribution of the $\ell = 4$ multipole is always negligible. This justifies the fact that we do not consider multipoles of higher order.

\vspace{.5cm}
\begin{figure}
\includegraphics[width = 1\columnwidth]{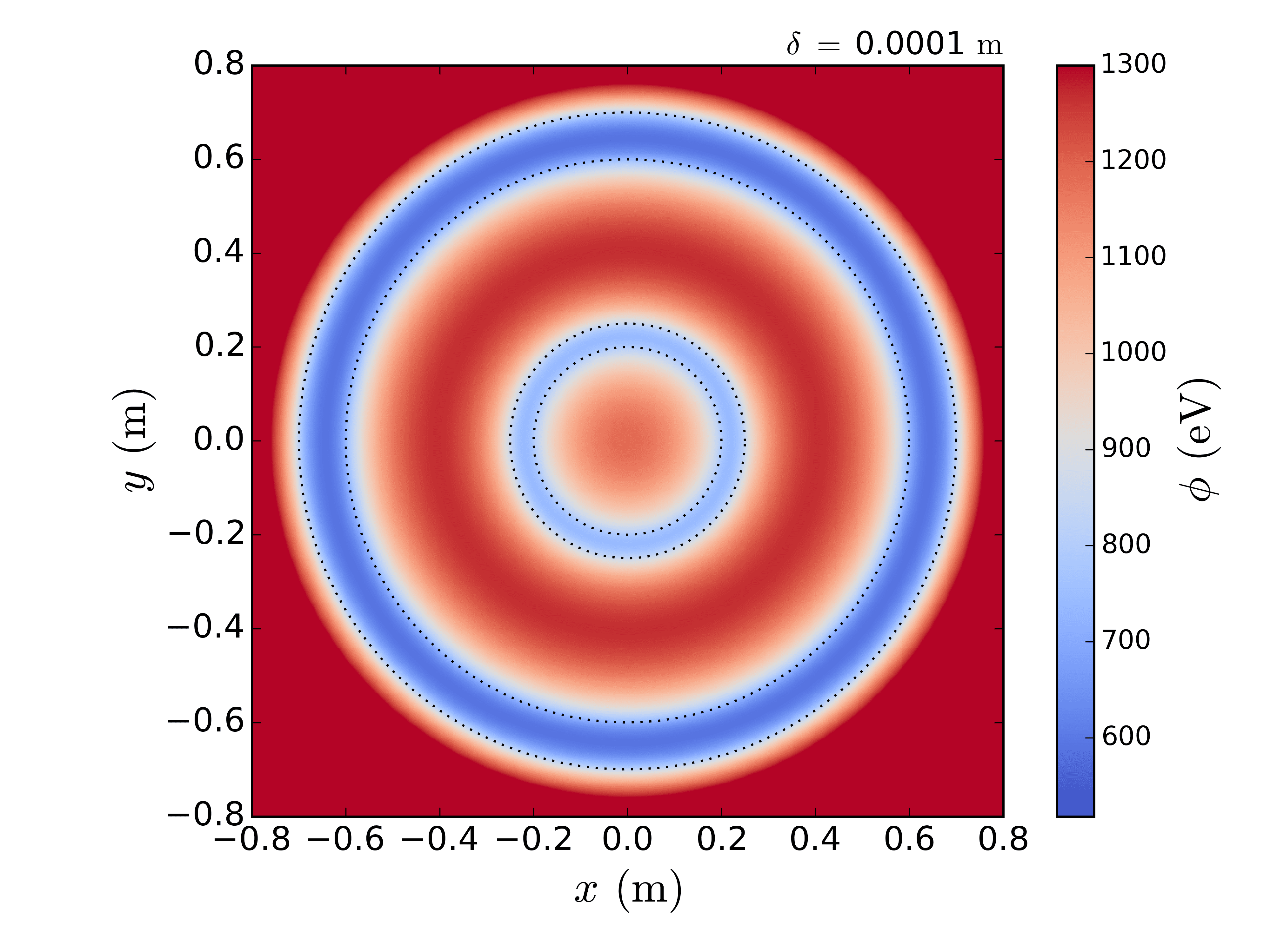}
\includegraphics[width = 1\columnwidth]{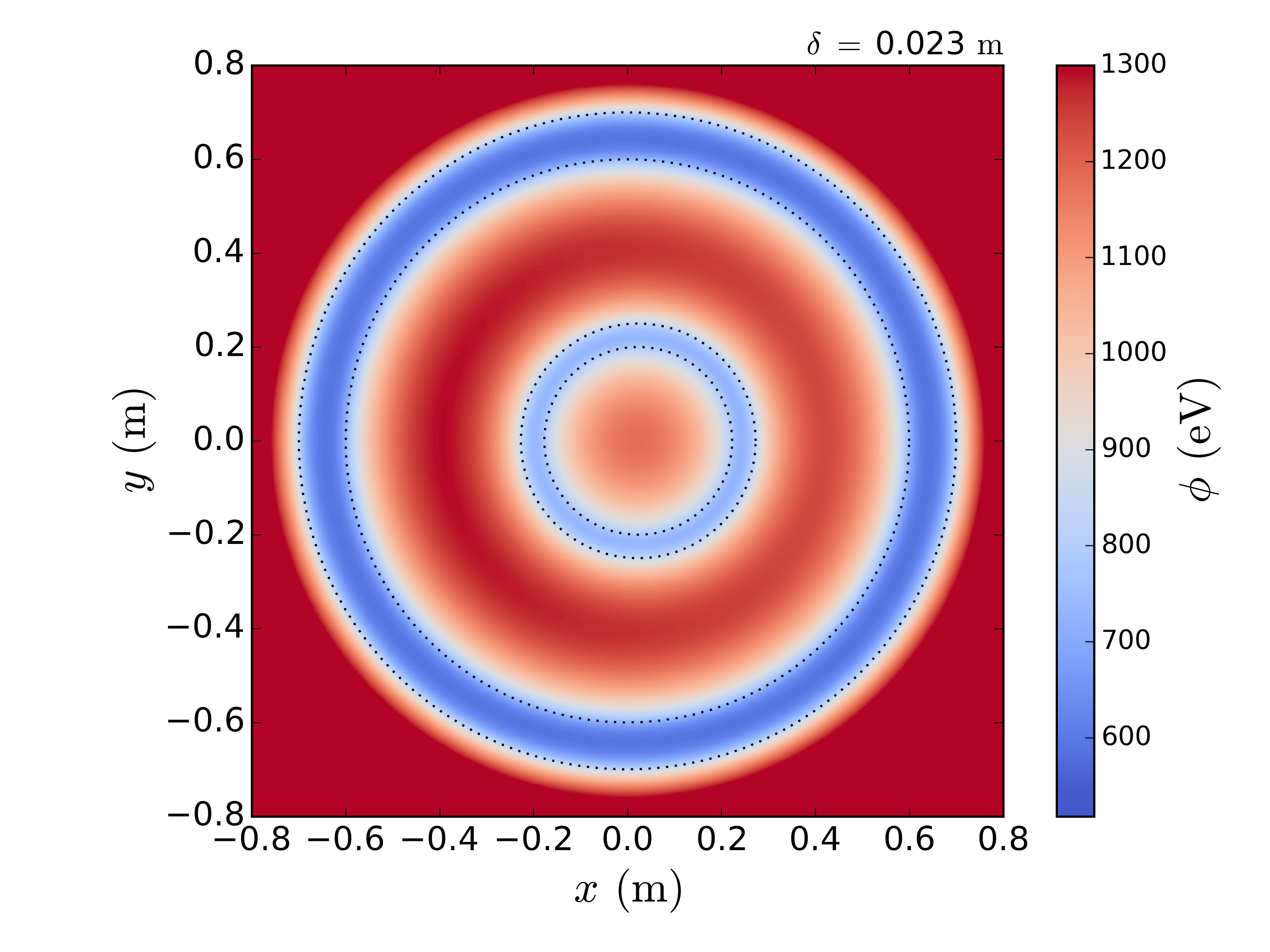}
\caption{Total field maps obtained by summing $\bar\phi$, $\psi_0$ and the multipoles for a set of displacement $\delta = 0.0001, \,0.023 \, {\rm m}$ from top to bottom. The dotted lines delimit the cylinders. The field is truncated at $1300\,{\rm eV}$ in this scale.}
\label{fig_maps}
\end{figure}

\begin{figure}
\includegraphics[width = 1.\columnwidth]{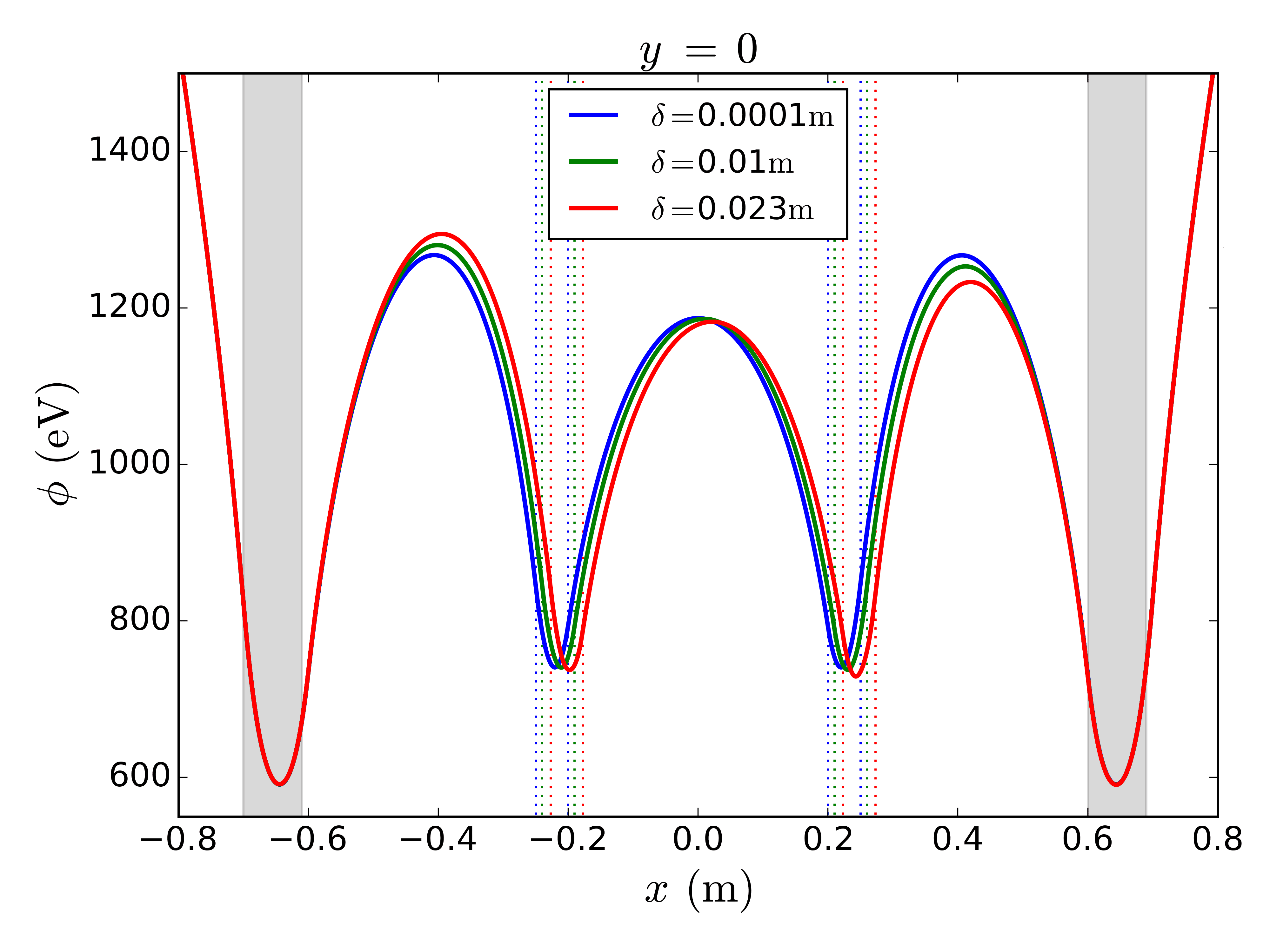}
\caption{Field profile slices for $y = 0$ for a set of displacement $\delta = 0.0001, \, 0.01, \,0.023 \, {\rm m}$. The shaded zones and the dotted lines delimit the cylinders.}
\label{fig_maps_slice}
\end{figure}

Now, from these multipoles we can reconstruct 2-dimensional maps of the field using Eq.~\eqref{e.decpsi}. Figure \ref{fig_maps} shows such maps for  different values of $\delta$. Figure \ref{fig_maps_slice} gives a clearer view of these maps showing slices of the field profile in the plane $y=0$. One can notice that asymmetry in the field appears along the axis of displacement. This is significant in the inter-cylinder space, where the field gets shrunk on the right side of the inner cylinder while expanding on the left. Similarly, the maximum of the field in the space enclosed by the inner cylinder departs from $x=0$. When integrating over the whole cylinder this will be responsible for a force on the inner cylinder.

\subsection{Accuracy of the approximation}\label{sec:approxchk}

When solving  Eqs.~(\ref{e.2cyl-hierarchy0Fin}-\ref{e.2cyl-hierarchyellFin}), we neglected the non-linear terms in $u_\ell$ -- $3^{\rm rd}$ in the r.h.s. of both Eqs.~\eqref{e.2cyl-hierarchy0Fin} and~\eqref{e.2cyl-hierarchyellFin}. Unfortunately, when evaluating them with the solution we have obtained, we notice that despite the suppression at high $r$ caused by the powers of $\frac{\delta}{r}$, they can dominate close to the inner cylinder. This occurs for multipoles of order $l \geqslant 2$.

To verify the impact of this terms, we solve again the multipole equations~(\ref{e.2cyl-hierarchy0Fin}-\ref{e.2cyl-hierarchyellFin}) taking into account the non-linear terms that we evaluate with the solution we first obtained by neglecting them. These terms involve a sum over all multipoles and we only keep terms up to $\ell = 4$ which is justified by the hierarchy of the multipoles observed on Fig.~\ref{fig_ul}. As expected, this procedure leaves the monopole and the dipole unchanged, whereas for higher multipoles there is a subsequent change in their relative magnitude while their global shape is conserved. This however has a limited impact on the total field and on the associated force as we will show that the monopole and the dipole are the dominant contributions to the force. The impact lessen for smaller displacement $\delta$. The multipole shown in Fig.~\ref{fig_ul} take into account these non-linear corrections.

\section{Two-dimensional cylindrical asymmetric configuration:  full numerical computation}\label{sec:fullnum}

We can also address the problem of the nested cylinders by a full numerical 2-dimensional simulation, that will not rely on the approximations of the previous section. We follow the same approach as Ref.~\cite{elder_chameleon_2016} that uses an iterative relaxation algorithm  which, from an initial guess, converges slowly to the solution. We apply it to the 2-dimensional chameleon equation
\begin{equation}
\frac{\partial^2\phi}{\partial x^2} + \frac{\partial^2\phi}{\partial y^2} = \frac{\beta}{{\rm M}_{\rm Pl}}\rho({x, y})- n \phi^{-(n+1)},
\end{equation}
which is  discretized over a Cartesian 2D mesh by Taylor expanding to get
\begin{equation}
\begin{split}
\frac{\phi_{i+1,j}-2\phi_{i,j}+\phi_{i-1,j}}{(\Delta x)^2} &+ \frac{\phi_{i,j+1}-2\phi_{i,j}+\phi_{i,j-1}}{(\Delta y)^2} \\
&= \frac{\beta}{{\rm M}_{\rm Pl}}\rho({x_i, y_j})- n\, (\phi_{i,j})^{-(n+1)},
\end{split}
\end{equation}
where $\phi_{i,j}$ denotes the field  in the cell $(i,j)$ of the mesh, $\Delta x$ and $\Delta y$ the resolutions of the mesh along the two axis. Here, we use a square mesh so that $\Delta x = \Delta y$. Then, starting from an initial guess we can iteratively redefine the field over the mesh as
\begin{equation}
\begin{split}
\Phi^{(k+1)}_{i,j} = &\frac{\phi^{(k)}_{i+1,j} + \phi^{(k)}_{i-1,j} + \phi^{(k)}_{i,j+1} + \phi^{(k)}_{i,j-1}}{4} \\
&- \frac{(\Delta x)^2}{4} \left[\frac{\beta}{{\rm M}_{\rm Pl}}\rho({x_i, y_j})- n\, (\phi^{(k)}_{i,j})^{-(n+1)}\right]
\end{split}
\end{equation}
where $k$ denotes the iteration. The process thus consists, at each iteration, in taking the mean value of the field on the 4 closest neighbors to which one subtracts $(\Delta x)^2$ times the second member of the equation evaluated with the current solution.  After enough iterations this converges to the solution as long as the resolution of the mesh is fine enough. Having a resolution a tenth smaller than the smallest Compton wavelength of the field in the considered set-up --  here $\lambda_{\rm c,in}$ -- is sufficient by inspection.

Nevertheless, due to the non linearity of the equation, instabilities can appear. To overcome them we use a under-relaxation process, by adding a part of the $k^{\rm th}$ solution in the redefinition the $k+1^{\rm th}$ as
\begin{equation}
\phi^{(k+1)}_{i,j} = (1-\omega) ~\phi^{(k)}_{i,j} + \omega ~\Phi^{(k+1)}_{i,j}
\end{equation}
where $\omega$ is the over-relaxation parameter that we take as $\omega = 0.9$ and $\Phi$ is defined by the previous equation.

In this method, due to the finish extent of the mesh, we must set boundary conditions at finite distance unlike the method used in the previous section. In our case, this requires the external cylinder to be thick enough for the field to reach the minimum of its potential, so that the internal field becomes screened. In our previous work~\cite{PhysRevD.100.084006}, we showed that for a wall to be safely screened, its thickness needs to be roughly larger than $100\, \lambda_{\rm c, wall}$. Here, for the parameters we consider, due to the limited computing resources, we have only been able to use a mesh allowing one to have an external cylinder of thickness $80\, \lambda_{\rm c, wall}$, which appears to be sufficient.

Note that we are also limited by the facts we need to have a large enough mesh to treat  the boundary conditions correctly and to have a precise enough mesh to model the small variations of the field that are more likely to happen inside the cylinders, which are the very quantity needed to evaluate the force. This limits us for exploring the chameleon parameter space, and makes this method complementary to the one presented in the previous section. This problem is less likely to be encountered in Ref.~\cite{elder_chameleon_2016} as it focused on the field variations in the vacuum gaps and thus could neglect all variations smaller than $\Delta x$, which anyway have a limited impact on the larger scale variations.

\subsection{Results}

\begin{figure}
\includegraphics[width = 1\columnwidth]{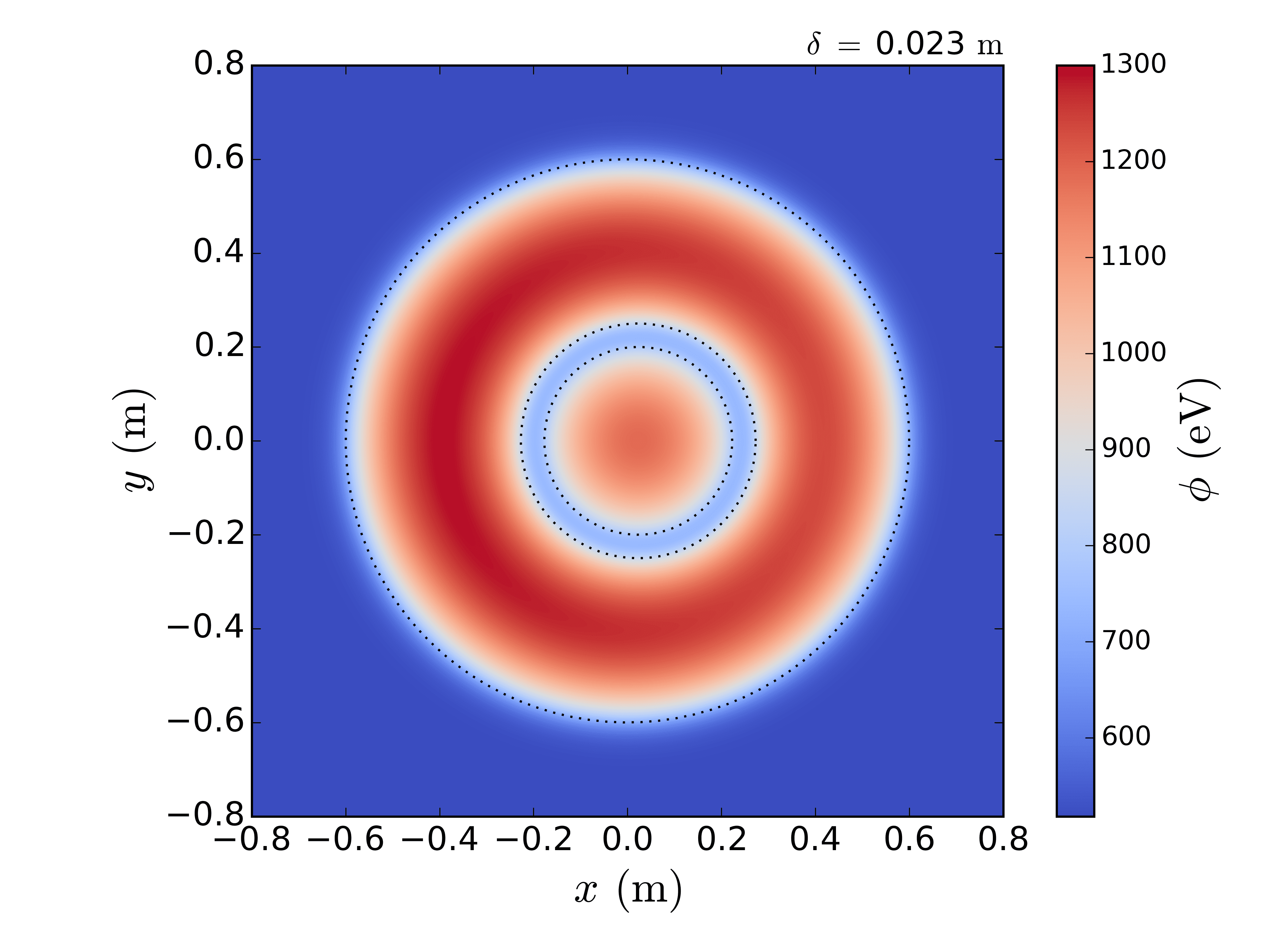}
\includegraphics[width = 1\columnwidth]{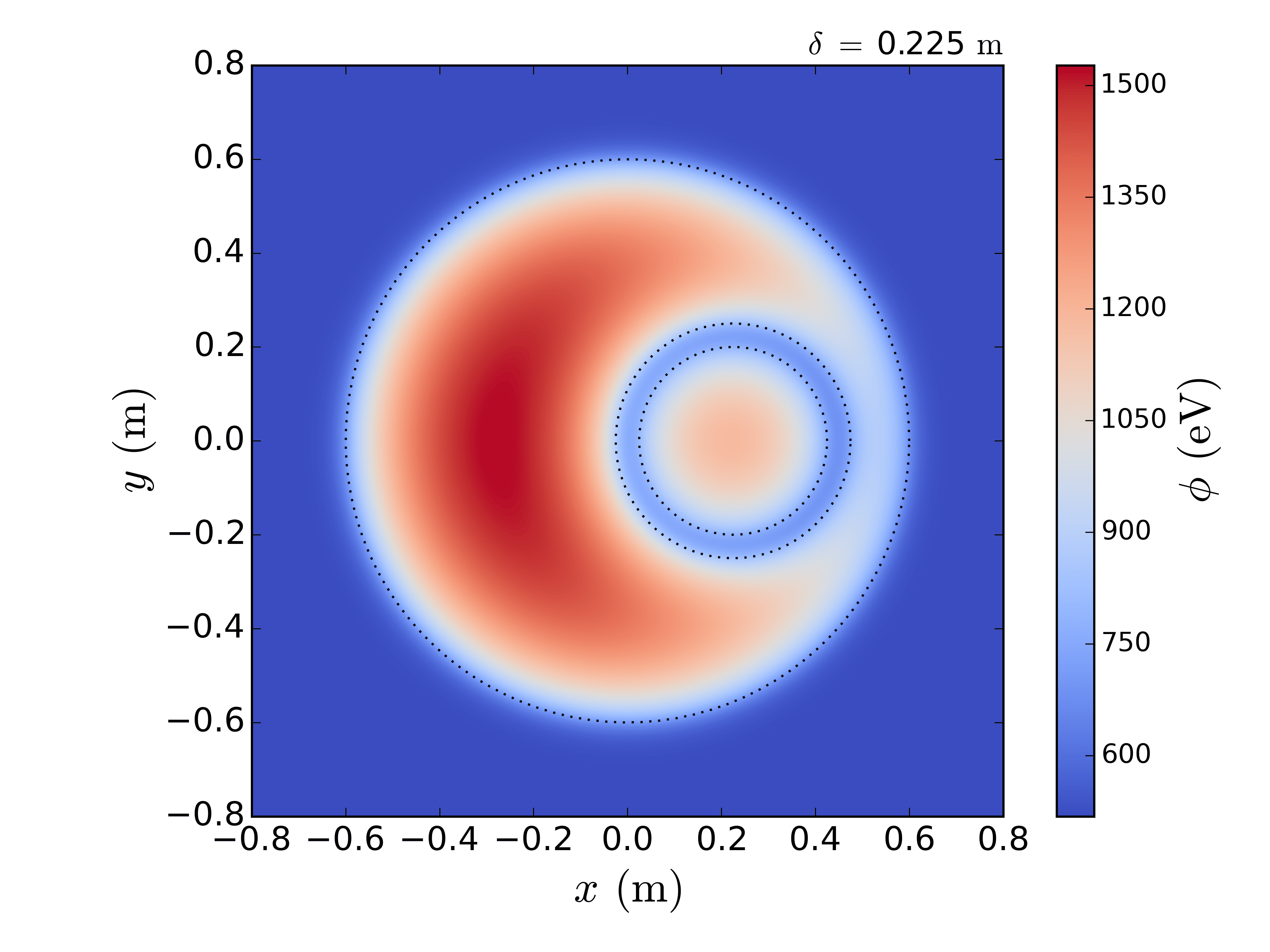}
\caption{Total field maps obtained by a full numerical simulation for a set of displacements $\delta = 0.023, \,0.225 \, {\rm m}$ from top to bottom. The dotted lines delimit the cylinders.}
\label{fig_maps2D}
\end{figure}

\begin{figure}
\includegraphics[width = 1.\columnwidth]{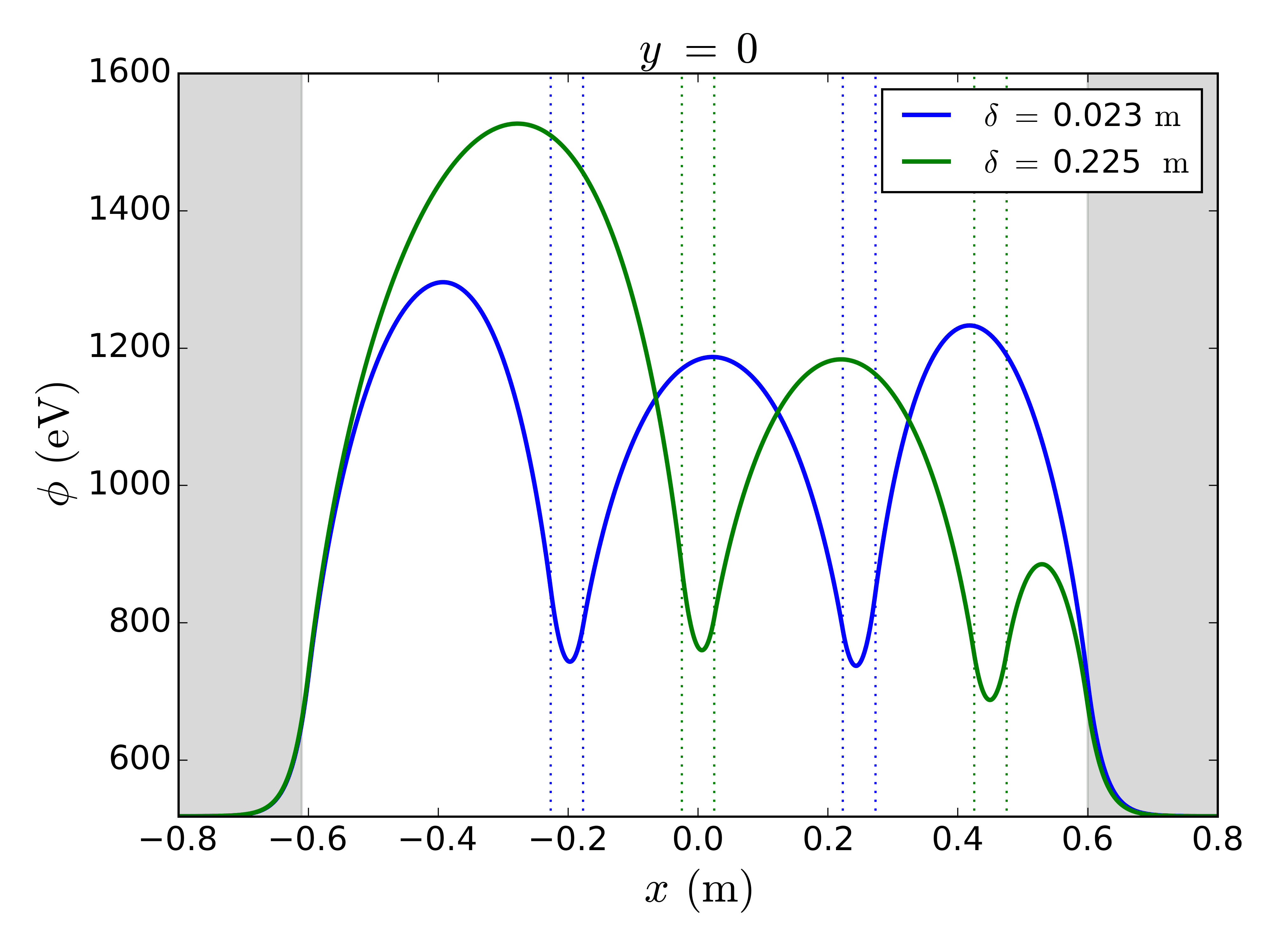}
\caption{Field profile slices in the plane $y = 0$ obtained by a full numerical simulation for a set of displacements $\delta = 0.023, \,0.225 \, {\rm m}$. The shaded zones and the dotted lines delimit the cylinders.}
\label{fig_maps_slice2D}
\end{figure}

The results of this methods are  displayed in Figs.~\ref{fig_maps2D} and~\ref{fig_maps_slice2D}. This method allows us to simulate larger displacements than the multipole method. The structure is faithful to the one observed in the previous section. We observe the different behaviors for $r > 0.6 \,{\rm m}$, due to the different ways of setting boundary conditions. The departure from cylindrical symmetry is more clearly apparent for large $\delta$, specially on both sides of the inner cylinder testifying of a more intense force.

\subsection{Comparison of the two methods}

\begin{figure}
\includegraphics[width = 1.\columnwidth]{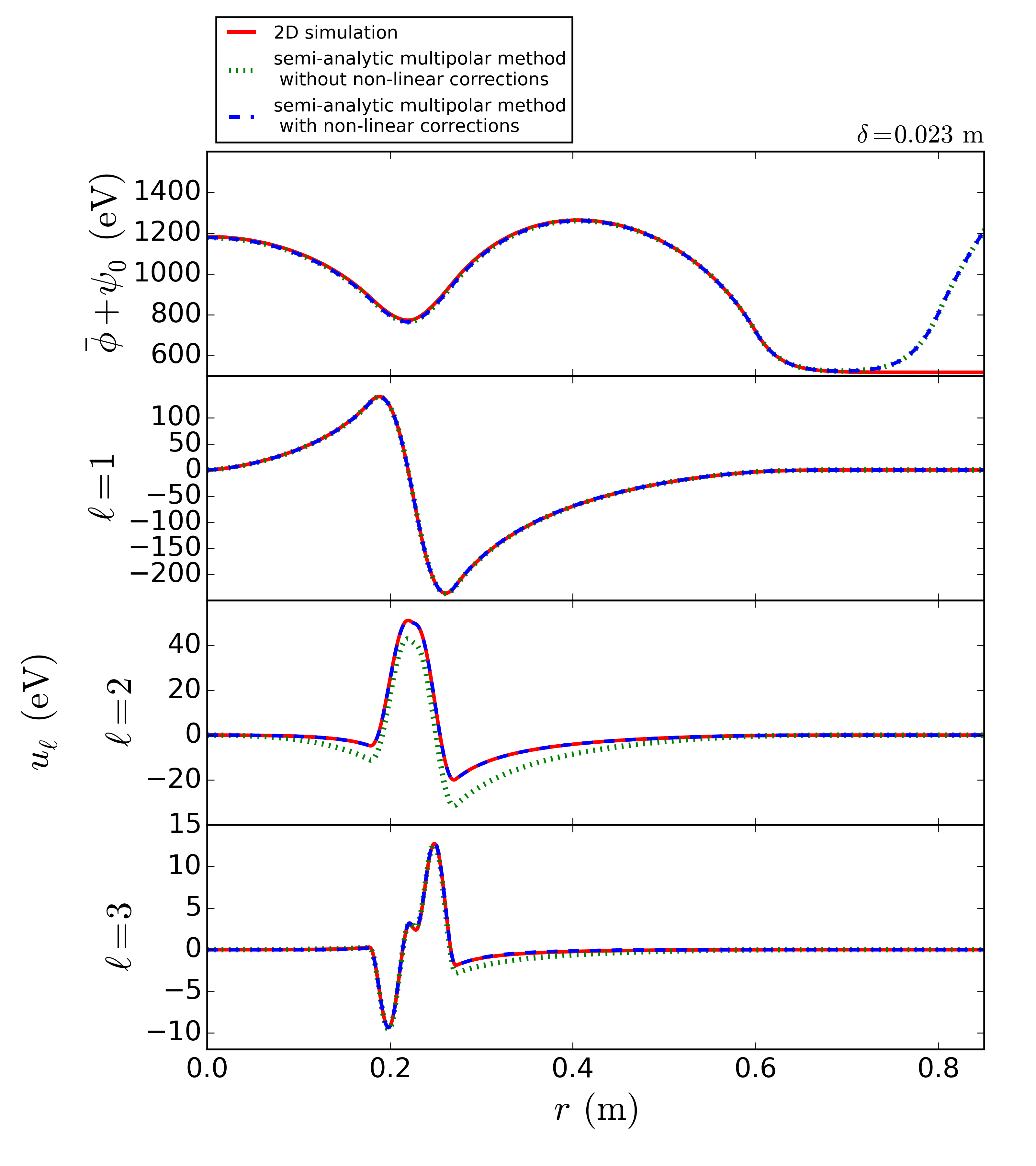}
\caption{Comparison of the multipoles computed by a full numerical method and the multipolar expansion method with and without considering terms non-linear in $u_{\ell}$ in Eq.~\eqref{e.2cyl-hierarchyellFin}.}
\label{fig_comp_ul}
\end{figure}

With the full numerical method we can treat any displacements that are larger than the resolution of the mesh. This overlaps with the previous semi-analytical method and enables us to compare them. To that end, we must increase the thickness of the external cylinder in the multipolar method. We are however limited by the numerical precision, we use a thickness of $0.2 \, {\rm m}$. Figure \ref{fig_comp_ul} compares the first multipoles computed in the previous section by considering or not the correction of the non-linear terms discussed in Sec.~\ref{sec:approxchk} to the multipoles extracted from the 2D simulation.

We observe that, as expected, the non-linear corrections have no impact on the monopole and the dipole. And the agreement is such that the largest difference between the multipoles obtain by the method amounts to less than a percent whether we considered non-linear corrections or not. For the modes $\ell=2$ and $\ell = 3$, without non-linear corrections, the multipolar expansion fails at reproducing the result of the 2D simulation, reaching differences in the multipoles that represent an error of 69\%. Fortunately, when we consider non-linear corrections this differences falls to respectively less than a percent for $\ell = 2$ and 9\% for $\ell = 3$. This is a very strong confirmation of the validity of the multipolar approximation. The difference of 9\% for $\ell = 3$ will be negligible when considering the force exerted on the inner cylinder, as we shall discuss.

\section{Force between cylinders}\label{sec:force}

Now, we have all the elements to study the force that the cylinders are experiencing when shifting the inner one by $\delta$.

\subsection{Definition of the force}

The force on the inner cylinder, is obtained by integrating the fifth force on the cylinder, hence
\begin{equation}
{\bf F} = - \frac{\beta}{{\rm M}_{ \rm Pl}}
 \int {\nabla}\phi {\rm d}m
= - \frac{\beta}{{\rm M}_{ \rm Pl}}  \rho_{\rm in} h {\bf\cal F}[\delta]
\end{equation}
with
\begin{equation}
  {\bf\cal F}[\delta]\equiv \int_0^{2\pi} {\rm d}\theta \int_{r_-(\theta)}^{r_+(\theta)}  \nabla\phi    r {\rm d}r.
\end{equation}
We denote ${{\bf F}_h} = \frac{{\bf F}}{h}$. ${{\bf F}_h}$ and ${\bf\cal F}$ only differ by a constant factor of $- \beta \rho_{\rm in}/{\rm M}_{ \rm Pl}$.

Since we assume a displacement along the $x$-axis, the $y$-components on two symmetric elements (i.e. on $\theta$ and $2\pi-\theta$) are equal and opposite so that
\begin{equation}
   {\cal F}_x= {\cal F}, \qquad  {\cal F}_y=0.
\end{equation}
It follows that
\begin{equation}\label{e.FF}
  {\bf\cal F}[\delta]\equiv \int_0^{2\pi} {\rm d}\theta \int_{r_-(\theta)}^{r_+(\theta)} \left[\cos\theta \partial_r \phi - \frac{\sin\theta}{r}\partial_\theta\phi\right]    r {\rm d}r.
\end{equation}
Replacing the multipolar expansion of the field we obtain
\begin{eqnarray}\label{e.FFgen}
  {\bf\cal F}[\delta] &=& \int_0^{2\pi} \cos\theta  \int_{r_-(\theta)}^{r_+(\theta)} [\bar\phi'+\psi_0']r \dd r \\
    &+&\sum_{\ell\not0} \int_0^{2\pi} \hbox{e}^{i\ell \theta} \int_{r_-(\theta)}^{r_+(\theta) } \sqrt{\frac{\delta}{r}} \left[\left(u'_\ell -\frac{u_\ell }{2 r}\right) \cos\theta \right.\nonumber\\
    &-&\left.i\ell \frac{u_\ell}{r} \sin\theta \right]\nonumber
    r {\rm d}r.
\end{eqnarray}
We see that $\bar\phi'+\psi_0']$ will contribute to all the multipoles of the force.

\subsection{Computation of the force}

Let us proceed with a series of approximations that will allow us to get to the full  generic expression of the force. Those approximations turn to be useful to understand the magnitude of the force.

\subsubsection{Test inner cylinder}

First, we consider that the inner cylinder as a test cylinder in the sense that its presence does not affect the scalar field profile inside the cavity. The latter is thus purely dictated by the outer cylinder, thus is axially symmetric and given by $\bar\phi(r)$ alone. It follows that the expression~(\ref{e.FF}) reduces to
\begin{equation}\label{e.FF1cyl}
  {\bf\cal F \rvert_{\bar\phi}}[\delta]\equiv \int_0^{2\pi} \cos\theta {\rm d}\theta \int_{r_-(\theta)}^{r_+(\theta)}  \bar\phi'(r)   r {\rm d}r.
\end{equation}
Now, since
$$
r_-(\theta) \equiv R+h_-(\theta) =  R + \delta\cos\theta -\frac{1}{2}\frac{\delta^2}{R}\sin^2\theta + \ldots
$$
and
$$
r_+(\theta) \equiv R+e +h_+(\theta) = R+e + \delta\cos\theta -\frac{1}{2}\frac{\delta^2}{R+e}\sin^2\theta + \ldots
$$
where the dots contain terms which are higher powers of $\sin^2\theta$, we split the integral  over $r$ as
$$
\int_R^{R+e} - \int_R^{r_-} + \int_{R+e}^{r_+}.
$$
Obviously, the first does not depend on $\theta$ and gives $0$ after angular integration. The other two  reduce to $\bar\phi'(R)R[r_(\theta)-R]$ and $\bar\phi'(R+e)(R+e)[r_+(\theta)-R-e]$. When integrating over $\theta$ only the linear term in $\delta$ survives so we get
\begin{equation}\label{e.FF1cylFin}
  {\bf\cal F\rvert_{\bar\phi,{\rm lin}}}[\delta] = \left[\bar\phi'(R+e)(R+e)-\bar\phi'(R)R \right]\frac{\delta}{2}.
\end{equation}
In this approximation we can get the force directly from our the results of our former work~\cite{PhysRevD.100.084006}. Even though we assumed staticity, we can write down the equation of motion for the inner cylinder as $m\ddot\delta=F$ so that
$$
\ddot\delta + \frac{\beta}{2\pi M_p}\left[\frac{\bar\phi'(R+e)(R+e)-\bar\phi'(R)R}{(2R+e)e} \right] \delta =0,
$$
i.e we expect a typical pulsation of order
\begin{equation}\label{e.defw}
 \omega^2 = \frac{\beta}{2\pi M_p}\left[\frac{\bar\phi'(R+e)(R+e)-\bar\phi'(R)R}{(2R+e)e} \right].
\end{equation}
Note that this does not assume that $\omega^2$ is positive. If the slope of ${\cal F}[\delta]$ is positive then the force destabilizes the system  and $\omega$ has to be thought as the inverse of a stability time.

\begin{table}[h!]
\caption{Magnitude of the force and of the associated (inverse) of the stability time defined in Eq.~(\ref{e.defw}) in the inner cylinder test approximation.}
\begin{tabular}{*4c}
\toprule
$\delta ({\rm m})$  & $|{{\bf F}_h} \rvert_{\bar\phi}|({\rm N.m^{-1}})$ &   $|{{\bf F}_h} \rvert_{\bar\phi,{\rm lin}}| ({\rm N.m^{-1}})$  & $|\omega|\,({\rm rad.s^{-1}})$ \\
\hline
$10^{-6}$ & $9.57~10^{-11}$ & $3.57~10^{-11}$ & $1.18~10^{-4}$\\
$10^{-4}$ & $9.57~10^{-9}$ & $3.57~10^{-9}$ & $1.18~10^{-4}$\\
$10^{-2}$   & $9.57~10^{-7}$ & $3.57~10^{-7}$ & $1.18~10^{-4}$\\
0.023  	& $2.20~10^{-6}$ & $8.21~10^{-7}$ & $1.18~10^{-4}$ \\
\bottomrule
\label{table_phibar}
\end{tabular}
\end{table}

Table.~\ref{table_phibar} summarizes the force for different $\delta$ by the integration of $\bar\phi$ through both Eqs.~({\ref{e.FF1cyl}-\ref{e.FF1cylFin}), with or without the linear approximation. Both methods reproduce the same order of magnitude. We conclude that the force is positive so that the fifth force destabilizes the system of cylinders.

\subsubsection{Inner cylinder with radial backreaction}

To go one step further, we consider the change of the profile of the field induced by the inner cylinder but neglect the $\ell\not=0$ modes so that $\psi_0(r)$ is taken as the symmetric configuration when $\delta=0$.  It follows that the expression~(\ref{e.FF}) reduces to
\begin{equation} \label{e.FF2}
  {\bf\cal F\rvert_{\bar\phi+\psi_0}}[\delta]\equiv \int_0^{2\pi} \cos\theta {\rm d}\theta \int_{r_-(\theta)}^{r_+(\theta)}  [\bar\phi'(r) +\psi_0'(r)]  r {\rm d}r.
\end{equation}
This lets us to a similar computation as the previous one with a modified profile
\begin{equation}\label{e.FF2cylFin}
  {\bf\cal F\rvert_{\bar\phi+\psi_0,{\rm lin}}}[\delta] = \left[(\bar\phi'+\psi_0')_{(R+e)}(R+e)- (\bar\phi'+\psi'_0)_R R \right]\frac{\delta}{2}.
\end{equation}

\begin{table}[h!]
\caption{Magnitude of the force and of the associated pulsation taking into the cylindrically symmetric back-reaction. To be compared to Table~\ref{table_phibar}. Note the change of sign in the force that shows the stabilizing effect of the monopole.}
\begin{tabular}{*4c}
\toprule
$\delta ({\rm m})$  & $|{{\bf F}_h} \rvert_{\bar\phi+\psi_0}| ({\rm N.m^{-1}})$ &   $|{{\bf F}_h} \rvert_{\bar\phi+\psi_0,{\rm lin}}| ({\rm N.m^{-1}})$  & $|\omega|\,({\rm rad.s^{-1}})$                        \\ \hline
$10^{-6}$ & $-2.71~10^{-9}$ & $-3.20~10^{-10}$ & $3.54~10^{-4}$ \\
$10^{-4}$ & $-2.80~10^{-7}$ & $-3.20~10^{-8}$ & $3.54~10^{-4}$ \\
$10^{-2}$   & $-2.18~10^{-5}$ & $-2.87~10^{-6}$ & $3.35~10^{-4}$\\
0.023  	& $-3.55~10^{-5}$ & $-5.76~10^{-6}$ & $3.13~10^{-4}$  \\
\bottomrule
\label{table_psi0}
\end{tabular}
\end{table}

Table \ref{table_psi0} contains the values of the force applied to the inner cylinder corrected by the back reaction contribution of $\psi_0$, again by integrating it with or without the linear approximation for the force. Now for all $\delta$, the force is negative and the linear approximation fails to give the correct force by one order of magnitude. It shows that the monopole induces a stabilizing force, as can actually be seen directly from Fig.~\ref{fig_psi0} on which it can clearly be seen that the gradient of the scalar field becomes positive.

\subsubsection{Generic case}

The general expression~(\ref{e.FFgen}) includes the sum
\begin{equation}
{\sum_\ell \int_0^{2\pi} \hbox{e}^{i\ell \theta} \int_{r_-(\theta)}^{r_+(\theta) }
     \sqrt{\frac{\delta}{r}} \left[\left(u'_\ell -\frac{u_\ell }{2 r}\right) \cos\theta - i\ell \frac{u_\ell}{r} \sin\theta \right]
    r {\rm d}r.}
\label{force_ul}
\end{equation}
Again in the small $\delta$ limit, this can be computed by splitting the integral over $r$ as
\begin{eqnarray*}
&& \cos\theta \int_{R}^{R+e} \sqrt{\frac{\delta}{r}} \left(u'_\ell(r) -\frac{u_\ell(r) }{2 r}\right) r \dd r \\
&& - i\ell \sin\theta  \int_{R}^{R+e} \sqrt{\frac{\delta}{r}} u_\ell(r)  \dd r \\
&& + \cos\theta \int_{R+e}^{R+e+h_+(\theta)} \sqrt{\frac{\delta}{r}} \left(u'_\ell(r) -\frac{u_\ell(r) }{2 r}\right) r \dd r \\
&& - i\ell \sin\theta \int_{R+e}^{R+e+h_+(\theta)} \sqrt{\frac{\delta}{r}} u_\ell(r)  \dd r \\
&& - \cos\theta \int_{R}^{R+h_-(\theta)} \sqrt{\frac{\delta}{r}} \left(u'_\ell(r) -\frac{u_\ell(r) }{2 r}\right) r \dd r \\
&&  + i\ell \sin\theta \int_{R}^{R+h_-(\theta)} \sqrt{\frac{\delta}{r}} u_\ell(r)  \dd r.
\end{eqnarray*}
Hence, ${\cal F}$ is obtained by integrating over $\theta$ the following expression
\begin{equation}
\begin{split}
& \hbox{e}^{i\ell \theta} \cos\theta \int_{R}^{R+e} \sqrt{\frac{\delta}{r}} \left(u'_\ell(r) -\frac{u_\ell(r) }{2 r}\right) r \dd r \\
& - i\ell ~\hbox{e}^{i\ell \theta}  \sin\theta  \int_{R}^{R+e} \sqrt{\frac{\delta}{r}} u_\ell(r)  \dd r \\
& + \hbox{e}^{i\ell \theta}  \cos\theta ~h_+(\theta) \left( \sqrt{\frac{\delta}{r}} \left(u'_\ell(r) -\frac{u_\ell(r) }{2 r}\right) r \right)_{r=R+e}\\
& - i\ell ~\hbox{e}^{i\ell \theta} \sin\theta h_+(\theta) \left(\sqrt{\frac{\delta}{r}} u_\ell(r)  \right)_{r=R+e}
\end{split}\nonumber
\end{equation}
\begin{equation}
\begin{split}
& - \hbox{e}^{i\ell \theta} \cos\theta ~h_-(\theta) \left( \sqrt{\frac{\delta}{r}} \left(u'_\ell(r) -\frac{u_\ell(r) }{2 r}\right) r \right)_{r=R} \\
&  +i\ell ~\hbox{e}^{i\ell \theta}  \sin\theta ~h_-(\theta)\left( \sqrt{\frac{\delta}{r}} u_\ell(r)  \right)_{r=R}.
\end{split}
\label{force_ul_lin}
\end{equation}
The first two terms have a contribution of $\ell=\pm1$ which scales as $\sqrt{\delta}$. The terms in $\delta\cos\theta$ in $h_\pm$ leads to terms linear in $\delta$ for $\ell=2$. Then higher multipoles arise from the shape $h_\pm(\theta)$. Basically we will have a series with terms scaling as $[\delta^2\sin^2(\theta)]^p$ each of which will involve multipoles up to $\ell=2p+1$ and each term is a higher power of $\delta^2$. This a good news since it better justifies the approximation scheme.

\begin{table}[h!]
\caption{Magnitude of the first multipoles of the force taking into account non-linearities (top) and in the linear approximation (bottom).}
\begin{tabular}{*5c}
\toprule
{} & \multicolumn{4}{c}{$|{{\bf F}_h} \rvert_{u_{\ell}+u_{-\ell}}| ({\rm N.m^{-1}})$}\\
$\delta \,({\rm m})$  & $\ell = 1$ & $\ell = 2$ & $\ell = 3$ & $\ell = 4$\\ \hline
$10^{-6}$ & $2.75~10^{-9}$ & $1.14~10^{-16}$ & -- & -- \\
$10^{-4}$ & $2.74~10^{-7}$ & $9.77~10^{-11}$ & $1.29~10^{-11}$ & -- \\
$10^{-2}$   & $2.11~10^{-5}$ & $8.71~10^{-7}$ & $1.11~10^{-7}$ & $7.40~10^{-9}$ \\
0.023  	& $3.12~10^{-5}$ & $4.27~10^{-6}$ & $5.42~10^{-7}$ & $3.65~10^{-8}$ \\
\bottomrule
\end{tabular}
\begin{tabular}{*5c}
\toprule
{} & \multicolumn{4}{c}{$|{{\bf F}_h} \rvert_{u_{\ell}+u_{-\ell},{\rm lin}}| ({\rm N.m^{-1}})$}\\
$\delta \,({\rm m})$  & $\ell = 1$ & $\ell = 2$ & $\ell = 3$ & $\ell = 4$\\ \hline
$10^{-6}$ & $2.74~10^{-9}$ & $5.06~10^{-17}$ & -- & -- \\
$10^{-4}$ & $2.74~10^{-7}$ & $5.06~10^{-11}$ & $1.16~10^{-15}$ & -- \\
$10^{-2}$   & $2.41~10^{-5}$ & $6.80~10^{-7}$ & $1.16~10^{-10}$ & $7.49~10^{-23}$ \\
0.023  	& $4.67~10^{-5}$ & $3.28~10^{-6}$ & $2.44~10^{-9}$ & $0$ \\
\bottomrule
\end{tabular}
\label{table_ul}
\end{table}

The expected tendency, deduced from our analytical analysis, that the contributions decrease with $\ell$ is numerically confirmed. Table~\ref{table_ul} shows the force computed for each multipole, as shown in Fig.~\ref{fig_ul}, up to $\ell=4$, i.e. the contribution to the force  resulting from the integration of Eq.~\eqref{force_ul} $|{{\bf F}_h} \rvert_{u_{\ell}+u_{-\ell}}|$ compared to the integration of Eq.~\eqref{force_ul_lin} $|{{\bf F}_h} \rvert_{u_{\ell}+u_{-\ell},{\rm lin}}| $ in the linear approximation. In both cases the multipole $\ell$ contains the contribution of $u_\ell$ and $u_{-\ell}$ to get a real-valued quantity. We observe, as expected, that the magnitudes of the multipoles decrease with higher $\ell$. This decrease is slower than what expected in the linear approximation of Eq.~\eqref{force_ul_lin}. For small $\delta$, we can consider that only the dipole contributes significantly to the total force. For larger $\delta$ the contributions are more balanced, but still, the multipoles with $\ell > 2$ can be neglected. In any case, the main contribution to the force are the monopole and dipole of the field and none can be neglected.

\subsection{Dependance of the total force on $\delta$}

\begin{figure}[]
\includegraphics[width = 1\columnwidth]{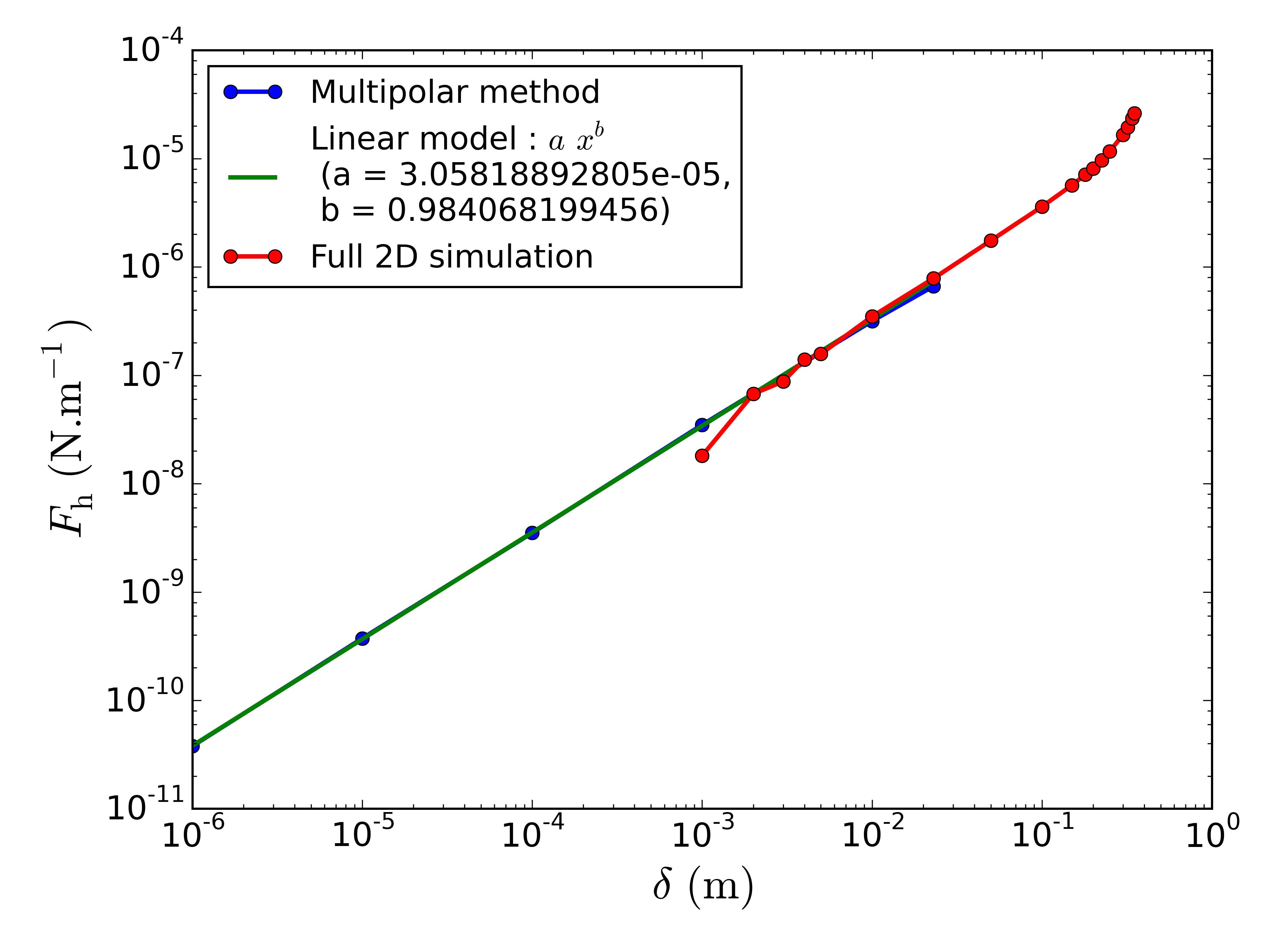}
\caption{Total force as a function $\delta$. The blue line refers to the result of the multipolar expansion method while the red line is the result of the full numerical simulation. The green line is a linear model fitted on the two first points.}
\label{fig_Fdelta}
\end{figure}

We can now gather all the different contributions and calculate the total force on the inner cylinder. Figure \ref{fig_Fdelta} depicts how it behaves with $\delta$. The sum on $\ell$ is truncated to $\ell = 2$ included. The force  is repulsive and linear in the displacement. The force obtained by both methods have been compared: there is an overlap for $\delta$ between $10^{-3}$ and $2.10^{-2} \,{\rm m}$, where both methods agree. At each limit of this interval, each method starts to show some  of its limits by departing from linearity. For the full numerical method, it is due to the fact that the mesh is too coarse compared to $\delta$. For the multipolar method, it is due to the fact that some higher non-linear terms we have not considered become non-negligible for large $\delta$. Nevertheless, both method are consistent and show the same global linear behavior and magnitude. The linearity of the force occur for $\delta \lesssim 10^{-1}\,{\rm m}$ and have linear stiffness of $k_h = - 3. 10^{-5} \, {\rm N.m^{-2}}$ -- assuming the convention ${\bf F} = - k\, {\boldsymbol  \delta}$, and $k_h = k/h$.

\begin{figure}[]
\includegraphics[width = 1\columnwidth]{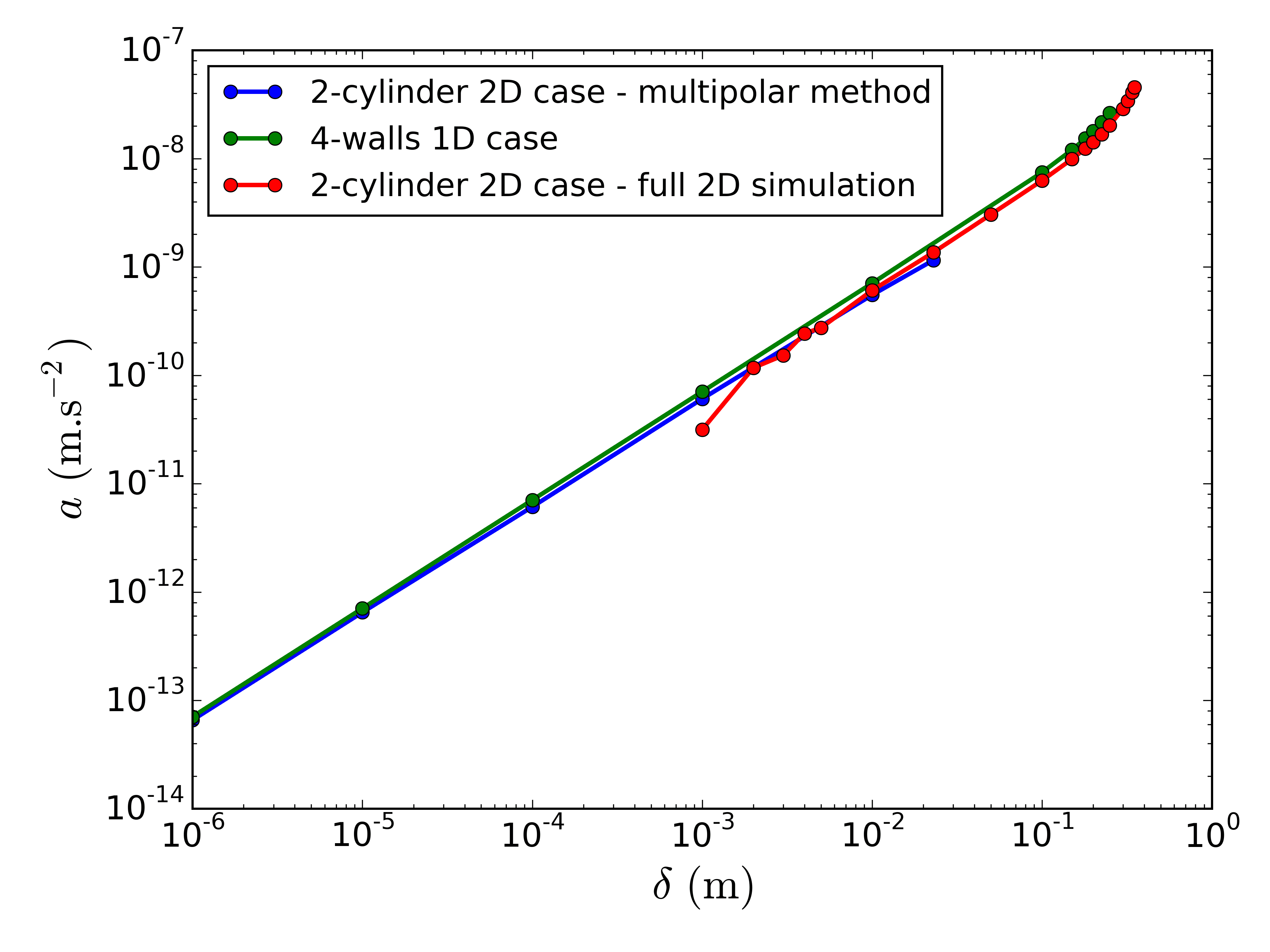}
\caption{Acceleration experienced by the inner cylinder of a 2-cylinders configuration computed by the multipolar expansion method blue)and the full numerical simulation (red). The green line represents the acceleration experienced by the two central wall of a 4-wall configuration, i.e. of the analog 1-dimensional problem.}
\label{fig_accdelta}
\end{figure}

These results can also be compared with the one-dimensional simulation of asymmetry from Sec.~\ref{sec:1D}. To be comparable to this cylindrical case, we consider a 4-wall configuration where the two internal walls move together. We compute the acceleration experienced by respectively the two walls and the internal cylinder. Figure \ref{fig_accdelta} shows that both cases are in excellent agreement. The linearity of the force occurs for the same range of $\delta$ and the departure from linearity for large $\delta$ are very much similar. The magnitude of the accelerations using both methods differ by a factor smaller than 2, so that the cylindrical geometry does not bring any major additional contribution to the force -- it even lowers it slightly.

\subsection{Total force variation with $\beta$ and $\lambda$}\label{sec:Fbetalambda}

Let us investigate the dependence of this fifth force on the chameleon parameters $\beta$ and $\Lambda$. We run the multipolar method for different parameters for $\delta = 10^{-6} \, {\rm m}$, and compare $k_h$ that we estimate as the linear slope of $F(\delta)$.

\begin{table}[h!]
\caption{Dependence of the slope per unit of length of the cylinder, $k_h=-F/(h\delta)$ , of the pressure with the parameters $\beta$ and $\Lambda$ of the chameleon model.}
\centering
{\tiny
\begin{tabular}{rc|ccccc}
        &  $k_h$    &     &   & $\Lambda$ &   &    \\
        & {\tiny N.m$^{-2}$}    & 0.4 & 1 & 3         & 5 & 10 \\ \hline
        & 0.01 & $6.72\,10^{-10}$  & $2.81\,10^{-10}$ & $2.31\,10^{-10}$ & $2.26\,10^{-10}$ & $2.24\,10^{-10}$ \\
        & 0.1  & $1.25\,10^{-6}$ & $5.13\,10^{-7}$ & $3.91\,10^{-8}$ & $3.01\,10^{-8}$ & $2.65\,10^{-8}$ \\
$\beta$ & 1    & $5.55\,10^{-6}$ & $3.78\,10^{-5}$ & $1.69\,10^{-4}$ & $3.67\,10^{-5}$ & $5.55\,10^{-6}$ \\
        & 4    & $4.37\,10^{-5}$ & $8.78\,10^{-5}$ & $8.85\,10^{-4}$ & $2.20\,10^{-3}$ & $1.57\,10^{-3}$ \\
        & 10   & -- & $3.03\,10^{-4}$ & $1.41\,10^{-3}$ & $4.46\,10^{-3}$ & $1.51\,10^{-3}$
\end{tabular}}
\label{tabkcham}
\end{table}

\begin{figure}[]
\includegraphics[width = 1\columnwidth]{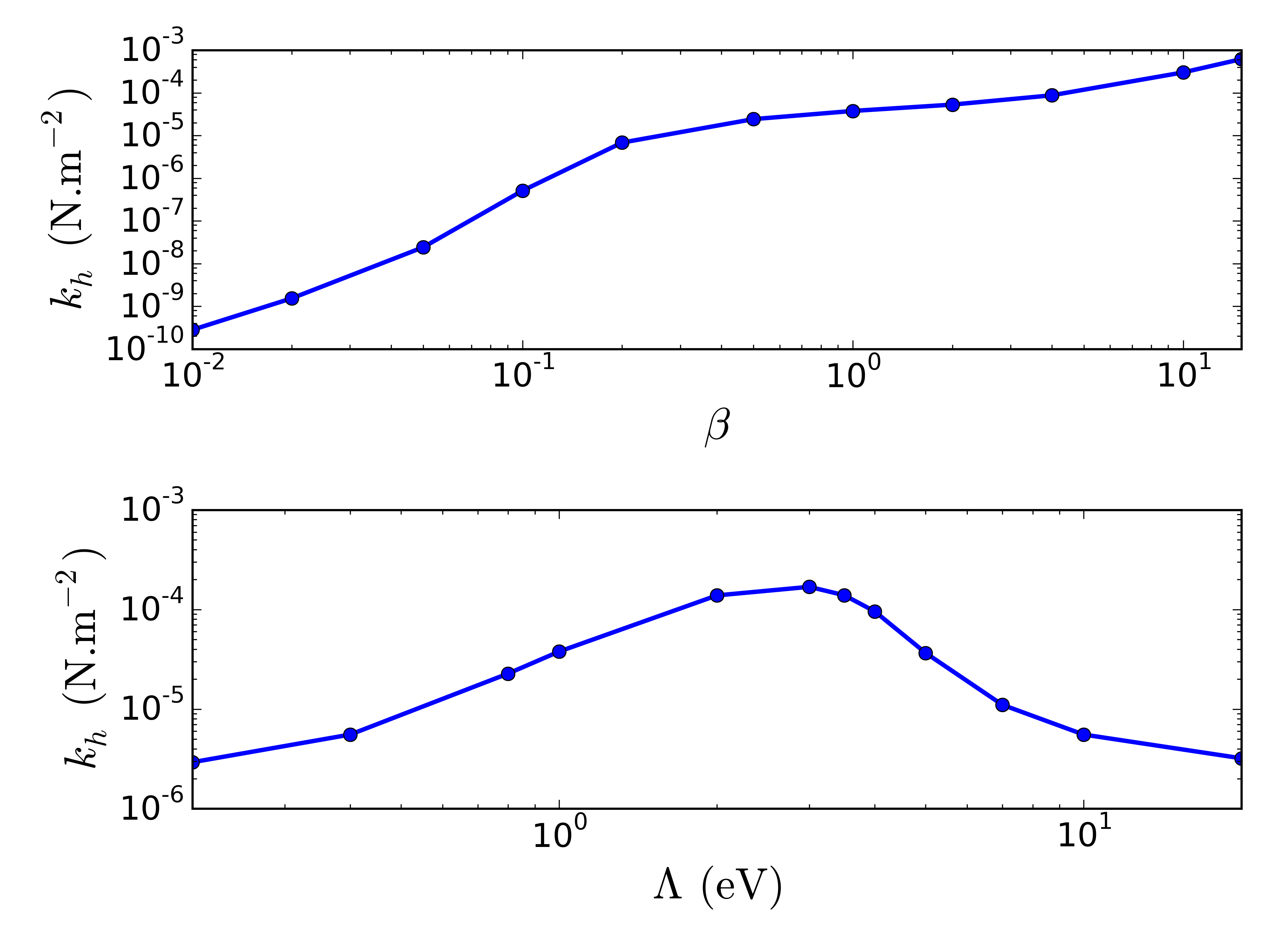}
\caption{Dependence of the chameleon stiffness $k_h$ to $\beta$ and $\Lambda$ for $\Lambda = 1 \,{\rm eV}$ and $\beta = 1$ respectively.}
\label{fig_k}
\end{figure}

Table~\ref{tabkcham} summarizes the values of $k_h$ obtained for different couples of parameters $(\beta,\Lambda)$. For each $k_h$, the sum of the multipole contribution is truncated at $\ell = 2$ as the next contribution are negligible.  Figure~\ref{fig_k} shows graphically its variation with $\beta$ for $\Lambda = 1 \,{\rm eV}$, and with $\Lambda$ for $\beta = 1$. The force increases with $\beta$ and exhibits a maximum along the $\Lambda$-axis, that flattens for small values of $\beta$. Notice that the behavior of the stiffness is similar for the 1-dimensional case of 4-walls. This behavior is interesting considering screening. The cylinders indeed tend to be screened for large $\beta$ and small $\Lambda$, as the Compton wavelength decreases. The behavior of the force shows that it can still be relevant even when the system of cylinders is screened -- $\beta$ large. This is promising as this could still lead to a detectable internal effect even when screening occurs, i.e. when  externally sourced effects are shielded, see Ref.~\cite{PhysRevD.100.084006}.
	
\subsection{Dependence on the geometry}

For now, we have fixed the  geometry with specific sizes of cylinders, gaps, and matter densities. Varying these parameters will indeed change the value of the force and its stiffness, as well as shifting the sensitivity curves displayed in Fig.~\ref{fig_k}.

 \subsubsection{Effect of the densities}

In most experiments, the vacuum density is much smaller than the one used in our analysis. Here, we estimate how this impacts the force by varying the density of the inter-cylinder vacuum. Figure~\ref{rhotestfig} shows the result for a displacement $\delta = 10^{-6} \, {\rm m}$ -- for higher $\delta$, the curve remains similar. The inter-cylinder vacuum density is expressed as a multiple of the cylinder density $\rho_{\rm in}$, which we keep fixed. So far we used $\rho_{\rm vac}/\rho_{\rm in} = 10^{-3}$.

\begin{figure}[]
\includegraphics[width = 1\columnwidth]{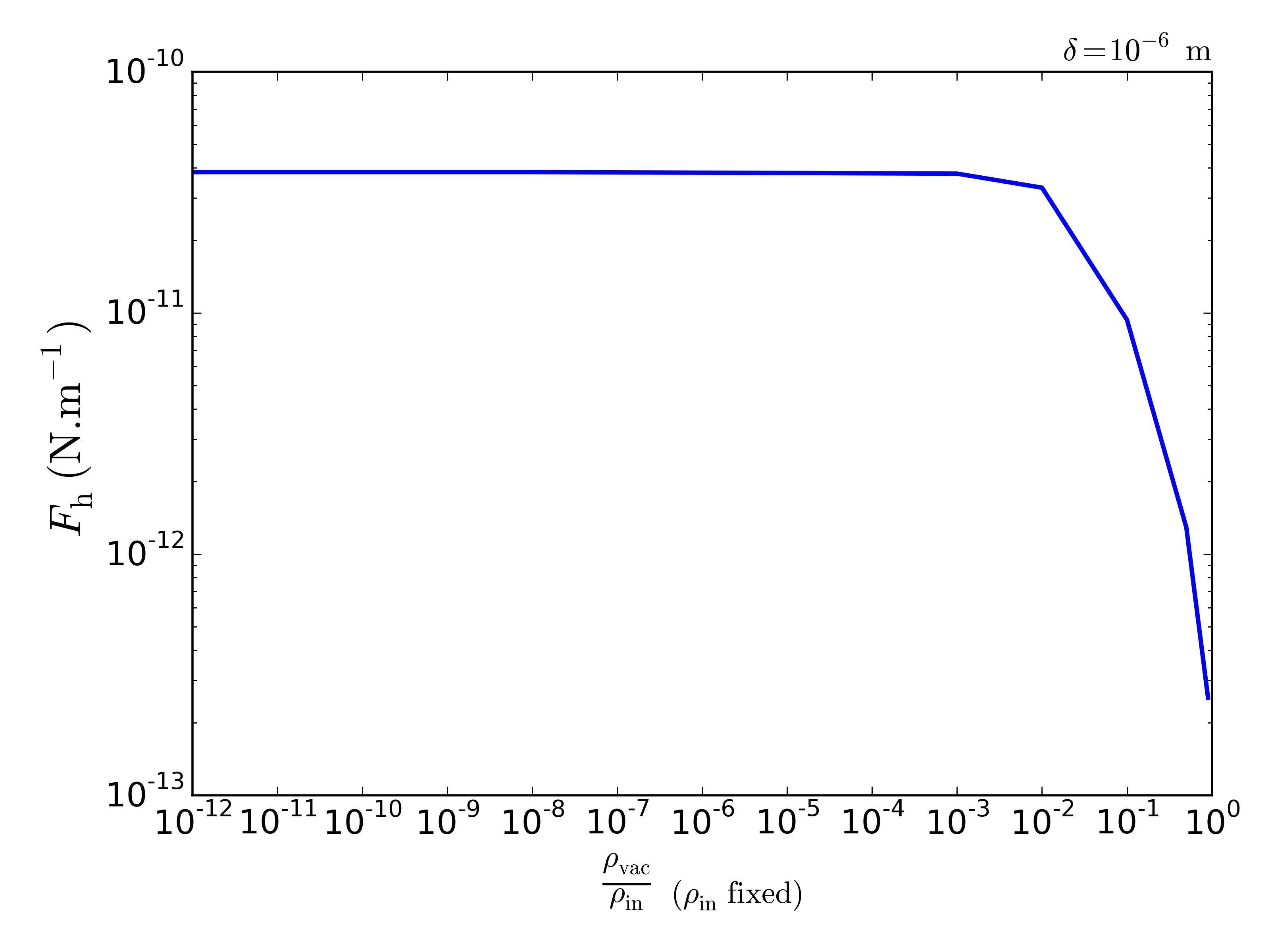}
\caption{Evolution of the force with the inter-cylinder vacuum density while the cylinder density is kept fixed, for $\delta = 10^{-6} \,{\rm m}$ and $n=2$, $\beta = 1$, $\Lambda = 1 \, {\rm eV}$.}
\label{rhotestfig}
\end{figure}

We observe that on the one hand, improving the vacuum quality leaves unchanged the magnitude of the force. This is due to the fact that the field is in fact unchanged in the inter-cylinder and exterior regions. When lowering $\rho_{\rm vac}$ the associated Compton wavelength stretches such that the field has less room to vary, but the associated minimum of the potential $\phi_*$ gets stretched at the same time. These two effects compensate so that the profile and the force remains unchanged. On the other hand, when worsening the vacuum quality the force gets exponentially suppressed. This occurs when the Compton wavelength associated to $\rho_{\rm vac}$ becomes of same order of magnitude as the inter-cylinder gap, as then the field has enough room to reach its minimum so that the previous argument is no longer valid. The force becomes null when the vacuum density equals the density of the cylinders. This is natural as, in this case, the system can be considered as a solid cylinder in which the field is flat and equal $\phi_*(\rho_{\rm in})$ deeply inside the cylinder at the level of where the inner cylinder was.
This confirms  that everything we obtained previously with $\rho_{\rm vac}/\rho_{\rm in} = 10^{-3}$ is directly transposable to case of a better vacuum quality.

\subsubsection{Scaling of the geometry}

Considering smaller scales in the geometry by reducing the sizes of the cylinders and the gaps would also affect the force. The scaling mentioned in Eq.~\eqref{e.KG-gen} should give us the answer to  this question. Indeed, it gives a correspondence between two geometries with constant matter densities, as long as the chameleon parameters are changed accordingly. This can be generalized to a scaling of the type
\begin{eqnarray}
&&{\bf x} \rightarrow {\bf x'} = \alpha_x{\bf x}, \nonumber \\
&&\Lambda \rightarrow \Lambda' = \alpha_\Lambda \Lambda,  \nonumber  \\
&&\beta \rightarrow \beta' = \alpha_\beta \beta,  \nonumber \\
&&\phi \rightarrow \phi' =  \alpha_\phi \phi  \nonumber  \\
&&\rho \rightarrow \rho' =  \alpha_\rho^3 \rho,
\end{eqnarray}
keeping the Planck mass unchanged. In order for the field equation to be unchanged, we need to impose that
\be\label{e.v0}
 \alpha_\phi=\alpha_\beta\alpha_\rho^3\alpha_x^2, \quad\hbox{and}\quad
 \alpha_\phi^{n+2} = \alpha_\Lambda^{4+n}\alpha_x^2.
\ee
Hence, Eq.~\eqref{e.KG-gen} corresponds to $\alpha_\rho=1$, $\alpha_\Lambda=\alpha_\phi=\alpha$, $\alpha_x=1/\alpha$ and $\alpha_\beta=\alpha^3$. It follows that the masses of the cylinders scale as $m\rightarrow m'= \alpha_\rho^3\alpha_x^3$ m. Since the force is given by ${\bf F}_h = -\frac{\beta}{{\rm M_{Pl}}} \frac{m}{h} \,{\nabla}_{{\boldsymbol x}} \phi$, it follows that it scales as
$$
{\bf F}\rightarrow{\bf F'} = \alpha_\beta \alpha_\rho^3 \alpha_x \alpha_\phi \,{\bf F},
$$
and that, given the constraint~(\ref{e.v0}) the profile of the field is obtained from a simple rescaling as the Klein-Gordon equation remains unchanged, up to a general conformal factor. Hence, in the particular case of Eq.~\eqref{e.KG-gen},  ${\bf F'} = \alpha^3 \,{\bf F}$. This tells us that small systems are more likely to provide detectable forces since shrinking all physical dimensions by a factor $\alpha$ (keeping the same materials; $\alpha_\rho=1$) would increase the force by a factor $\alpha^3$. On the other hand, this corresponds to another theory as $\beta$ has also been changed. It follows that the dependence of the force on $\beta$ and $\Lambda$ is impacted accordingly so that the curves of Sec.~\ref{sec:Fbetalambda} should be shifted along the $\beta$- and $\Lambda$-axis in a way consistent with the above scaling relations. All these scalings have been checked using our simulations.

\section{Conclusions}

This article investigated the fifth force that arises on the detector of a gravity experiment, in the case of chameleon models. As the profile of the scalar field is affected by the local matter density, this  requires to determine solutions of the Klein-Gordon equation inside the instrument. To that goal, we modeled the accelerometer in the simplest way as two nested cylinders. We then extended our previous work~\cite{PhysRevD.100.084006} to take into account the fact that the cylinders may move, violating the axial symmetry, and hence creating a non-vanishing fifth force on the cylinders.

The computation of this force requires full numerical simulations but we estimated its magnitude and dependence on the geometry and the parameters of the model by first assuming that the cylinders are infinite. In such a situation, the Newton force between the two cylinders vanishes exactly. First, we considered an analog 1-dimensional model with 2 parallel walls containing a third wall that can move from its central position. Then, we explored the case of 2 infinite nested cylinders. We developed a semi-analytic method based on a multipolar expansion of the field. It allowed us to solve the Klein-Gordon equation iteratively. While the hierarchy of equations for the multipoles is a coupled system due to the non-linearity of the chameleon model, we showed that they decoupled for small displacement.  We thus solved these equation numerically, first in the linear approximation and then with the first non-linear term, and compared them with the profiles obtained from a full numerical simulation using a finite difference relaxation method. The two approaches are complementary and agree perfectly inside their common domain of applicability.

In all the cases studied, 1- or 2-dimensional, the force is linear in the displacement, as long at it is small compared to the radius of the cylinders. The fifth force is repulsive so that it does not stabilize the system by restoring the symmetry. Interestingly, the accelerations induced by this force in 1 or 2 dimensions are in very good agreement, testifying that there is no significant effect created by the cylindrical geometry. Then, we studied the dependence of this force on the chameleon parameters. We mainly showed that the force was increasing with $\beta$ leading to the conclusion that one could expect detectable effects even when the cylinders are screened. We exhibited some scaling relations between the geometry and the parameters of the model and  explored the sensitivity of the force to geometrical parameters. Two features have been explored: (1) we showed that the force was  constant regardless of the magnitude of the density in the vacuum of the inter-cylinder gaps as long as this density is small enough, i.e. the Compton wavelength of the field in vacuum is smaller than the sizes of the gaps. This makes all our results valid for realistic densities of vacuum. Finally (2) we showed that reducing the size of the cylinders simultaneously would affect the force in such a way that dividing them by a factor $\alpha$ would multiply the force by a factor $\alpha^3$, leading to forces more likely to be detectable for smaller system.

While this analysis gives a first insight on the effect of a chameleon fifth force on a space detector with a geometry close to the MICROSCOPE accelerometer, it is still simplified. First it assumes infinite cylinders. Indeed, with finite cylinders one expects edge effects which would require full 3-dimensional simulations. Besides, while the Newtonian force is strictly zero for 2 infinite nested cylinders, it will be non-vanishing for finite cylinders. This study  allows us to control such simulations in the limits $h/R\gg 1$. Then, we assume that the configuration of cylinders is static. While this is fine to compute the fifth force, it may not be adapted for a dynamical analysis. Such an analysis would require to study the relaxation of the field when the inner cylinder is moving and would challenge  the hypothesis of a frozen field.  Nevertheless, our formalism paves the way to study the effects of a chameleon fifth force on the detector of gravity experiments such as the MICROSCOPE mission \cite{artsuivant}. 

\section*{Acknowledgments}
We thank Manuel Rodrigues, Gilles M\'etris and Pierre Touboul for useful discussions and technical information about the MICROSCOPE instrument. We thank Arno Vantieghem for useful and friendly discussion about numerical aspects. We thank the members of the MICROSCOPE Science Working Group for allowing us to start this project and encouraging us to pursue it. We acknowledge the financial support of CNES through the APR program (``GMscope+'' project). MPB is supported by a CNES/ONERA PhD grant. This work uses technical details of the T-SAGE instrument, installed on the CNES-ESA-ONERA-CNRS-OCA-DLR-ZARM MICROSCOPE mission. This work is supported in part by the EU Horizon 2020 research and innovation program under the Marie-Sklodowska grant No. 690575. This article is based upon work related to the COST Action CA15117 (CANTATA) supported by COST (European Cooperation in Science and Technology).

\appendix

\section{Computation of $I_\ell(r)$}

To obtain eq.~\eqref{e.2cyl-hierarchy0b}, we need to compute integrals of $e^{i\ell \theta}  \Xi$ that contains terms like
$$
I_\ell(r) = \int\frac{\dd\theta}{2\pi} H[r-f(\theta)]\hbox{e}^{i\ell \theta}
$$
where $H$ is the Heaviside function and $f$ stands for $r_-$ or $r_+$.

At constant r, the equation $r=f(\theta)$ has then 2 opposite solutions in $\theta$ as $f(\theta)$ is the polar equation of a circle of radius $R$ displaced of $\delta$. These solutions exists only when $r\in[R-\delta,R+\delta]$ and are given by
\begin{equation}\label{e.compVarT}
\cos\vartheta(r)= \frac{r^2+\delta^2-R^2}{2\delta r},
\end{equation}
for which we keep only the positive root, the second being $-\vartheta(r)$. Then it is clear that $H[r-f(\theta)]=1$ for $\theta\in[-\vartheta(r),\vartheta(r)]$ so that
$$
I_\ell(r) = \int_{\vartheta(r)}^{-\vartheta(r)}\frac{\dd\theta}{2\pi}\hbox{e}^{i\ell \theta}
$$
and thus
$$
I_\ell(r) = - \frac{\sin\ell\vartheta(r)}{\pi\ell}.
$$
It follows that
\begin{eqnarray}\label{e.compXi}
&&\int \Xi[r;r_-(\theta), r_+(\theta)] \hbox{e}^{-i\ell\theta}\frac{{\rm d}\theta}{2\pi} = \nonumber\\
&&\qquad\qquad\qquad\qquad \frac{\sin\ell\vartheta_{+}(r)}{\pi\ell}- \frac{\sin\ell\vartheta_{-}(r)}{\pi\ell}
\end{eqnarray}
from which we deduce that
\begin{eqnarray}\label{e.compXi}
&&\int \Xi[r;r_-(\theta), r_+(\theta)] \frac{{\rm d}\theta}{2\pi} =  \frac{\vartheta_{+}(r)-\vartheta_{-}(r)}{\pi}.
\end{eqnarray}

\bibliographystyle{ieeetr}
\bibliography{Bib}

\end{document}